\documentclass[twocolumn,10pt,cleanfoot]{asme2ej}

\usepackage{graphicx,subfig}
\usepackage{amsmath,amsfonts,amssymb,amsbsy,amscd,amsgen}
\usepackage{enumitem}

\newcommand{\id}{\mathrm d}
\usepackage{bm}

\newcommand{\funcspace}{\mathcal U}
\newcommand{\sigalg}{\mathcal B}
\newcommand{\Pmeas}{\mu}
\newcommand{\solmap}[1]{S^{\, #1}}

\newcommand{\subs}[2]{#1_{\raisebox{-3pt}{$\scriptstyle {#2}$}}}
\newcommand{\obs}{f}
\newcommand{\attr}{\mathcal A}
\newcommand{\sech}{\mbox{sech}}

\newtheorem{defn}{Definition}

\title{Extreme Events: Mechanisms and Prediction}

\author{Mohammad Farazmand
    \affiliation{
        Department of Mechanical Engineering\\
        Massachusetts Institute of Technology\\
        Cambridge, MA 02139, USA\\
    Email: mfaraz@mit.edu
    }   
}

\author{Themistoklis P. Sapsis 
    \affiliation{Department of Mechanical Engineering\\
        Massachusetts Institute of Technology\\
        Cambridge, MA 02139, USA\\
    Email: sapsis@mit.edu
    }
}

\begin{document}

\maketitle    

\begin{abstract}
{\it Extreme events, such as rogue waves, earthquakes and stock market crashes, 
occur spontaneously in many dynamical systems. 
Because of their usually adverse consequences, quantification, prediction and mitigation of extreme 
events are highly desirable. Here, we review several aspects of extreme events in phenomena described by 
high-dimensional, chaotic dynamical systems. 
We specially focus on two pressing aspects of the problem: 
(i) Mechanisms underlying the formation of extreme events and 
(ii) Real-time prediction of extreme events. 
For each aspect, we explore methods relying on models, data or both. 
We discuss the strengths and limitations of each approach as well as possible future research directions. 
}
\end{abstract}


\section{Introduction}
Extreme events are observed in a variety of natural and engineering systems.
Examples include oceanic rogue waves~\cite{dysthe08,donelan2017}, 
extreme weather patterns~\cite{Ropelewski87,easterling2000,moy2002,scheffer2008}, 
earthquakes~\cite{usgs} and shocks in power grids~\cite{Crucitti2004,fang2012}.
These events are associated with abrupt changes in the state of the system and often
cause unfortunate humanitarian, environmental and financial impacts. As such, the prediction
and mitigation of extreme events are highly desired.

There are several outstanding challenges in dealing with extreme events.
These events often arise spontaneously with little to no apparent early warning signs. 
This renders their early prediction from direct observations a
particularly difficult task~\cite{scheffer2009,ghil2011,dakos2012}. In certain problems, such as earthquakes, 
reliable mathematical models capable of predicting the extreme events are not available yet~\cite{ben2012}. 

In other areas, such as weather prediction where more advanced models are in hand, accurate predictions 
require detailed knowledge of the present state of the system which is usually unavailable. 
The partial knowledge of the current state together with the chaotic nature of the system leads to uncertainty
in the future predictions. These uncertainties are particularly significant during the extreme episodes~\cite{murphy2004,scarrott2012,mohamad2015}. 

In addition, models of complex systems are usually tuned using data assimilation techniques. 
This involves selecting the model parameters so that its predictions match the existing empirical 
data. The effectiveness of data assimilation, however, is limited when 
it comes to rare extreme events due to the scarcity of 
observation data corresponding to these events~\cite{doucet2001,majda2012_book,vanden2013,altwegg2017}. 

These challenges to modeling and prediction of extreme events remain 
largely outstanding. The purpose of the present article is to review some 
of these challenges and to present the recent developments toward their resolution. 

\begin{figure}[h]
\centering
\includegraphics[width=.4\textwidth]{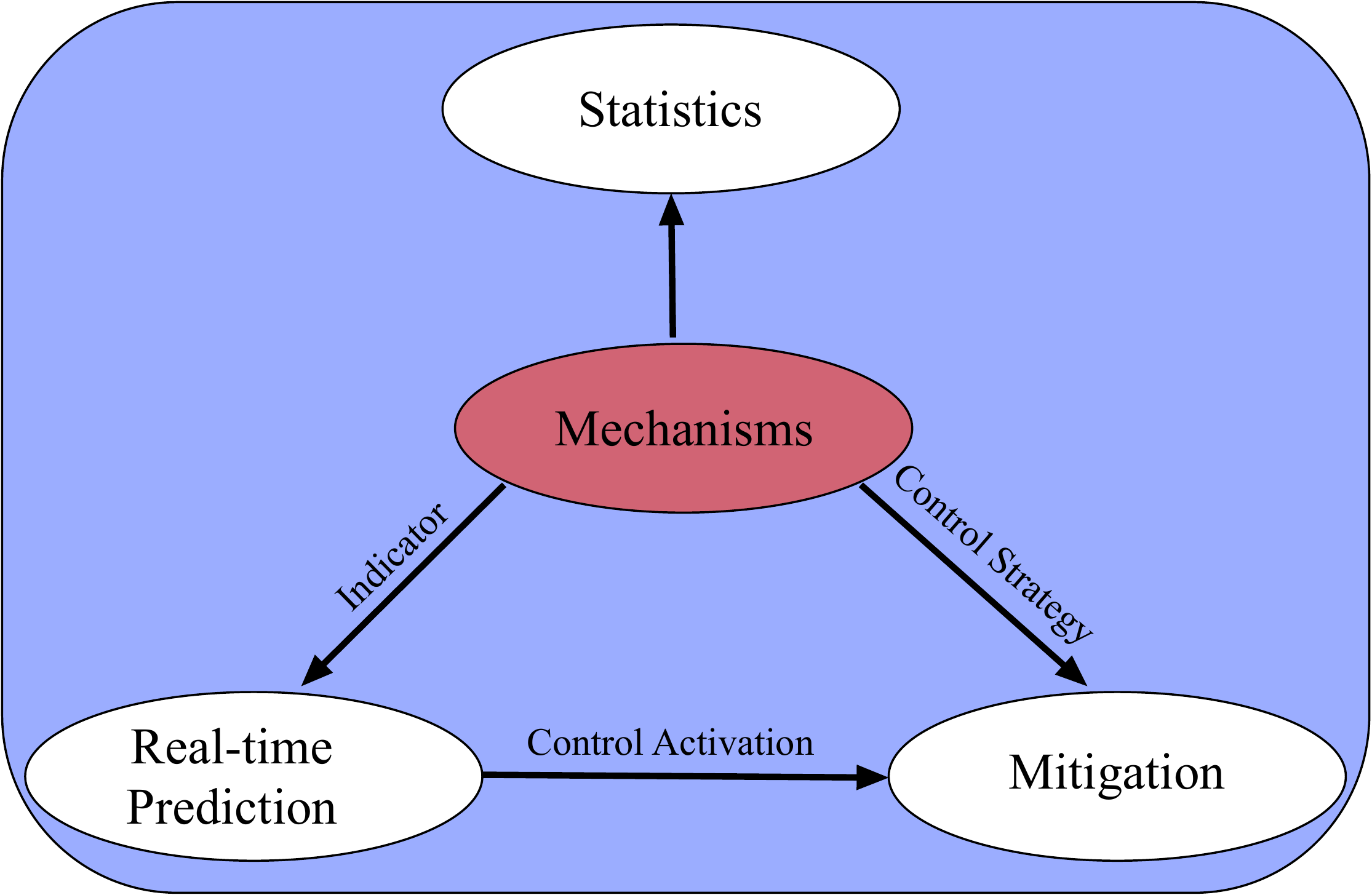}
\caption{The study of extreme events consists mainly of four components.}
\label{fig:ExtEventChart}
\end{figure}

The analysis of extreme events can be divided into four components as
illustrated in figure~\ref{fig:ExtEventChart}: Mechanisms, Prediction, Mitigation and Statistics. 
Below, we briefly discuss each of these components. 
\begin{enumerate}[label=(\roman*), wide, labelwidth=!]
\item Mechanisms: The mechanisms that trigger the extreme events are the primary focus of this article. 
Consider an evolving system that is known from the time series of its observables to produce extreme events. 
We are interested in understanding the conditions that underly the extreme events and trigger their formation.
Even when the governing equations of the system are known, it is often a difficult task to
deduce the mechanism underlying the extreme events. This is due to the well-known fact that even seemingly 
simple governing equations can generate very complex chaotic dynamics. The task of deducing the
behavior of solutions from the governing equations becomes specially daunting when the system consists of 
many interacting  degrees of freedom which give rise to a high-dimensional and complex attractor. 

In section~\ref{sec:geom}, we review a number of methods that unravel the extreme event mechanisms. 
These methods have been developed to analyze specific classes of dynamical systems. For instance, the 
multiscale method discussed in section~\ref{sec:slowfast} only applies to systems whose degrees of freedom can be separated 
into the so-called slow and fast variables. Even when such a slow-fast decomposition is available, 
computing the corresponding slow manifold and its stable and unstable manifolds become 
quickly prohibitive as the dimension of the system increases.

As a result, a more general mathematical framework is needed that is applicable to a broader range of 
dynamical systems and at the same time can leverage the ever growing computational resources. We explore
such a general framework in section~\ref{sec:probe}.

\item Real-time prediction: Most undesirable aspects of extreme events 
can often be avoided if the events are predicted in advance. 
For instance, if we can predict severe earthquakes a few hours in advance, many 
lives will be saved by evacuation of endangered zones. 
As a result, their real-time prediction is perhaps the most exigent aspect of extreme phenomena. 

Real-time prediction requires measurable observables that 
contain early warning signs of upcoming extreme events. We refer to such observable as indicators of extreme events. 
Reliable indicators of extreme events must have low rates of false-positive and false-negative predictions. A false positive
refers to the case where the indicator incorrectly predicts an upcoming extreme event. Conversely, a false negative 
refers to the case where the indicator fails to predict an actual extreme event. Knowing the mechanisms that trigger the extreme events does not necessarily enable their prediction. 
However, as we show in section~\ref{sec:prediction}, even partial knowledge of these mechanisms may lead
to the discovery of reliable indicators of extreme events. 

Another important aspect of extreme event prediction is the confidence in the predictions.
The sensitivity to initial conditions leads to an inherent uncertainty in chaotic systems even when 
the system model is deterministic~\cite{alligood1996,hirsch2012}. Such uncertainties permeate the prediction of extreme events.
As a result, the predictions have to be made in a probabilistic sense where the uncertainties in the
predictions are properly quantified (see section~\ref{sec:cond_stat}).

\item Mitigation: Can we control a system so as to suppress the formation of extreme events? 
This if of course beyond reach in many natural systems such as ocean waves and extreme weather patterns. 
However, in certain engineered systems, such as power grids, one can in principle design control strategies to
avoid the formation of extreme events~\cite{turitsyn2011,susuki2012,belk2016}. To this end, knowing the mechanisms that trigger the extreme events is 
crucial as it informs the design of the control strategy. The real-time prediction of the extreme events, on the other hand,
informs the optimal time for the activation of the control strategy (see figure~\ref{fig:ExtEventChart}).

The mitigation of extreme events within a dynamical systems framework has only recently been 
examined~\cite{ott2013,galuzio2014,Lai2014,Lai2015,bialonski2015,Joo2017}. The research in this direction has been limited to 
mitigation in simplified models by introducing arbitrary perturbations that nudge the system away from the extreme events.
However, a systematic study involving controllability and observability of extreme events in
the sense of control theory is missing. 

\item Statistics: The statistical study of extreme events attempts to answer questions regarding the frequency and probability of occurrence of
extreme events from a large sample. Such statistical questions are perhaps the most intensely studied aspect of extreme events due to their 
applications in finance, insurance industry and risk management~\cite{morgan1990,wilmott2007,McNeil15,Longin17}. 
In this article, we will limit our discuss of the statistics to this section and refer the interested reader to 
the cited literature on the topic.

Two major frameworks for quantifying the extreme statistics of stochastic processes are
the \emph{extreme value theory} and the \emph{large deviation theory}. 
The extreme value theory studies the probability distribution of the random variable 
$M_n=\max\{ X_1,X_2,\cdots, X_n\}$ where $X_1,X_2,\cdots$ is a sequence of
random variables~\cite{deHaan2007}. The main objective in extreme value theory
is to determine the possible limiting distributions of $M_n$ as $n$ tends to infinity. 
In particular, the Fisher--Tippett--Gnedenko theorem (also known as the extremal types theorem) states that, if
$\{X_i\}_{i\geq 1}$ is a sequence of independent and identically distributed (i.i.d) random variables
then the limiting distribution of $M_n$ can only converge to three possible distributions
and provides explicit formula for these distributions~\cite{frechet1927,fisher1928,Gnedenko43}. 
This is a significant result since the extreme statistics of the random
variable can be deduced even when no extreme events have actually been observed. 
In many practical cases, however, the random variables are not independent. Therefore, the 
more recent work in extreme value theory has been focused on relaxing the independence 
assumption~\cite{Watson54,Loynes65,Leadbetter1974,leadbetter1983,hsing1988,leadbetter1989,chernick1991,Freitas2008,Freitas2015}. 
For an extensive review of extreme value theory in the context of dynamical systems, we refer to a recent book by 
Lucarini et al.~\cite{lucarini2016}.

Another prominent framework for the statistical analysis of 
extreme events is the large deviation theory which is concerned with
the tail distribution of random variables. The tails of the probability distributions
contain the extreme values a random variable can take, hence the name large deviations. 
The large deviations were first analyzed by Cram\`er~\cite{cramer1938} who studied the
decay of the tail distribution of the empirical means $Z_n=\sum_{i=1}^nX_i/n$ for $n\gg 1$ where $\{X_i\}_{i\geq 1}$ 
is a sequence of i.i.d random variables. Later, Donsker and Varadhan~\cite{varadhan1975a,varadhan1975b,varadhan1976,varadhan1983}
generalized the large deviation results to apply them to Markov processes. 
The current scope of the large deviation theory is quite broad 
and is applied to quantifying heavy tailed statistics in a variety of deterministic
and stochastic dynamical systems. We refer the interested reader to the articles by Varadhan~\cite{varadhan2008} and Touchette~\cite{touchette2009,touchette2011} for a historical review of 
large deviation theory and its applications. 
\end{enumerate}

The four aspects of extreme events mentioned above are intertwined. However, the discovery of mechanisms 
that give rise to extreme events resides in the heart of the problem (see figure~\ref{fig:ExtEventChart}). 
For instance, even partial knowledge of the mechanisms that trigger the extreme events may lead to 
the discovery of indicators that facilitate their data-driven prediction (see section~\ref{sec:pred_app}). In addition, 
once we know what mechanisms trigger the extreme events, we can make informed choices about 
the control strategies towards avoiding them. To this end, the real-time prediction of upcoming extreme events
informs the time the control strategy should be activated. Knowledge of the mechanisms of the
extreme events can also help improve the statistical estimates regarding their likelihood and frequency. 

As a result, the main focus of this article will be on the first aspect of extreme events, i.e., the mechanisms. 
We will also discuss some aspect of the real-time prediction of the extremes, specially 
the quantification of the reliability of the indicators of extreme events. 
In section~\ref{sec:prelim}, we introduce the general set-up and notation.
Section~\ref{sec:geom} reviews some well-known mechanisms for extreme event formation in 
deterministic and stochastic dynamical systems. 
In section~\ref{sec:probe}, we review a variational method for 
discovering the mechanisms of extreme events and illustrate its application with two examples: 
intermittent turbulent energy dissipation and rogue ocean waves. 
In section~\ref{sec:prediction}, we discuss reliable indicators of extreme events
for their real-time prediction. Section~\ref{sec:concl} contains our concluding remarks. 

\section{Setup and notation}\label{sec:prelim}

In this section, we lay out the setup of the problem that allows for a dynamical systems
framework for extreme event analysis. We consider systems that are governed by an initial value problem
of the form
\begin{subequations}\label{eq:masterEq}
\begin{equation}
\partial_t u = N(u), 
\end{equation}
\begin{equation}
u(x,0)=u_0(x),\quad \forall x\in\Omega,
\end{equation}
\end{subequations}
where the state $u(t) \triangleq u(\cdot, t)\in \funcspace$ belongs to an appropriate function space $\funcspace$
for all times $t\geq 0$. The initial state of the system is specified by $u_0: \Omega \to \mathbb R^d$,
where $\Omega\subseteq \mathbb R^d$ and $d\in\mathbb N$. The operator $N$ is a potentially nonlinear
operator that is provided by the physics.
The PDE~\eqref{eq:masterEq} should also be supplied with appropriate boundary conditions $u|_{\partial\Omega}$ 
where $\partial\Omega$ denotes the boundary of $\Omega$.
\begin{figure*}
\centering
\includegraphics[width=.9\textwidth]{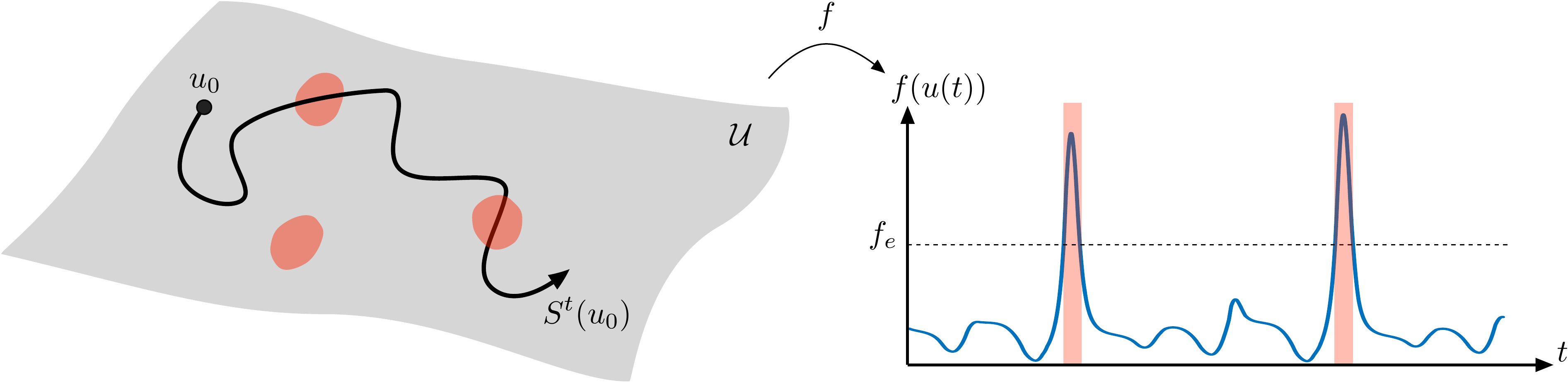}
\caption{Geometry of the state space $\funcspace$ and the time history of the observable $\obs$. As the trajectory passes through
the extreme event set $E_\obs(\obs_e)$-- marked in red- a burst in the observable time series appears.}
\label{fig:bursting_geom}
\end{figure*}

System~\eqref{eq:masterEq} generates a solution map 
\begin{align}
\solmap{t} : & \funcspace\to \funcspace\nonumber\\ 
& u_0\mapsto u(t)
\end{align}
that maps the initial state $u_0$ to its image $u(t)$ at a later time $t$. The solution map has the semi-group property, i.e., 
$\solmap{0}(u_0)=u_0$ and $\solmap{t+s}(u_0)=\solmap{t}(\solmap{s}(u_0))=\solmap{s}(\solmap{t}(u_0))$ for all $u_0\in\funcspace$.

We equip the space $\funcspace$ with further structure. In particular, we assume that $(\funcspace,\sigalg,\Pmeas)$
is a probability space and that the probability measure $\Pmeas$ is $\solmap{t}$-invariant. We refer to 
a measurable function $\obs:\funcspace \to \mathbb R$ as an observable. Note that for an observable $\obs$,
$X_t=\obs\circ \solmap{t}$ is a continuous stochastic process whose realizations are made by 
choosing an initial condition $u_0$ drawn in a fashion compatible with the 
probability measure $\Pmeas$. 

The observable $f$ is a quantity whose statistics and dynamical evolution is of interest. 
For instance, in the water wave problem considered in section~\ref{sec:wave} below, the observable is the wave height. In
meteorology, the observable of interest could be temperature or precipitation. 
Here, we are in particular interested in the extreme values of the 
observable $\obs$. In practice, the extreme values are
often defined by setting a threshold $\obs_e$. 
The observable values that are larger than this threshold constitute 
an extreme event. This motivates the following definition of extreme events.
\begin{figure*}
\centering
\includegraphics[width=.9\textwidth]{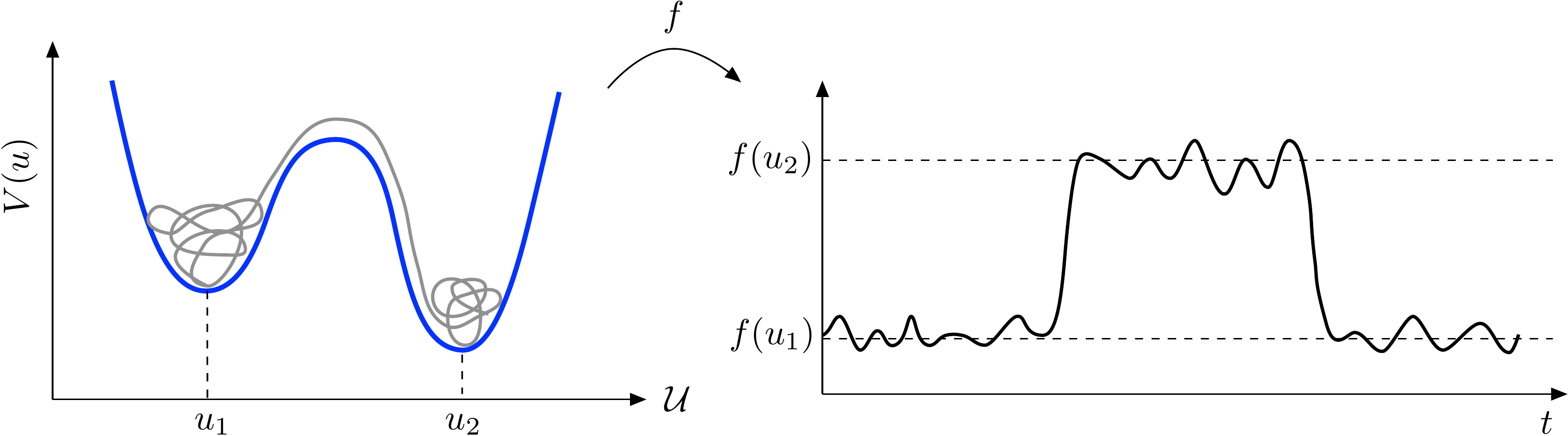}
\caption{Rare transitions between two stable states made possible through noise.}
\label{fig:transition_geom}
\end{figure*}

\begin{defn}[Extreme Events]
For an observable $\obs:\funcspace\to \mathbb R$, the extreme event set $E(\obs_e)$, corresponding to the 
prescribed extreme event threshold $\obs_e\in\mathbb R$, 
is given by
\begin{align}
E_\obs(\obs_e) & =\{u\in\funcspace : \obs(u)>\obs_e \}\nonumber\\
& = f^{-1}\big((f_e,\infty)\big).
\end{align}
\label{def:ee}
\end{defn}

The extreme event sets within the state space $\funcspace$ are depicted in figure~\ref{fig:bursting_geom}. 
As the system trajectory $\solmap{t}(u_0)$ passes through the extreme event set $E_\obs(\obs_e)$, 
the time series of the observable $\obs$ exhibit a sudden burst. In this figure, the 
extreme event set is depicted as a collection of patches, but in principle this set 
can have an extremely complex geometry. 

In certain problems, the extreme events may correspond to unusually small values of the observable $\obs$. 
In that case, Definition~\ref{def:ee} is still operative by studying the observable $-\obs$ instead of $\obs$.
A third type of rare events (which are not necessarily extreme) is the rare transition between long-lived states
(see figure~\ref{fig:transition_geom}). 
In this case, the system evolves for long times around a particular state before it is 
suddenly ejected to the neighborhood of a different state around which the system evolves 
for a long time before being ejected again~\cite{eyring1935,evans1935,laidler1983,truhlar1996,vanden2006,Metzner2006,TPT_ARPC}. 
Although such rare transitions do not necessarily fall under Definition~\ref{def:ee}, we return to them 
in section~\ref{sec:tpt} and review the mechanisms that cause the transitions. 

Finally, we point out that, although the governing equations~\eqref{eq:masterEq} are formulated as a partial differential equation (PDE),
we will also consider systems that are described by a set of ordinary differential equations (ODEs), $\dot u =N(u)$, 
where $u(t)\in\mathbb R^n$ denotes the state of the system at time $t$. This ODE could also arise from a finite-dimensional approximation of a
PDE model as is common in numerical discretization of PDEs.

\section{Routes to extreme events}\label{sec:geom}

There are certain classes of dynamical systems exhibiting extreme events
for which the mechanisms that trigger these events are well-understood. 
In this section, we review three such systems and discuss the underlying
mechanisms of extreme events in them. 

\subsection{Multiscale systems}\label{sec:slowfast}
An interesting type of extreme events appear in
slow-fast dynamical systems where the motion is separated into 
distinct timescales. 
The extreme events in such systems appear as
bursts when a system trajectory is dominated by the fast timescales of the
system. The early work on this subject was motivated by the observation of
relaxation oscillations in electrical circuits~\cite{vanderpol1926,vanderpol1934,benoit1983}.
Later, slow-fast dynamics found applications
in a wide range of problems such as chemical reactions~\cite{field1974,gillespie1977,connors1990,Koper1991,Arneodo1995}, 
excitable systems (e.g., neural networks)~\cite{ermentrout1986,rinzel1987,izhikevich2000,Guck02,ansmann2013,karnatak2014,saha2017}, 
 extreme weather patterns~\cite{Latif2009,dijkstra2013,roberts2016}, and dynamics of finite-size particles in fluid flows ~\cite{hallersapsis10,sapsis1}.

\begin{figure}
\centering
\includegraphics[width=.4\textwidth]{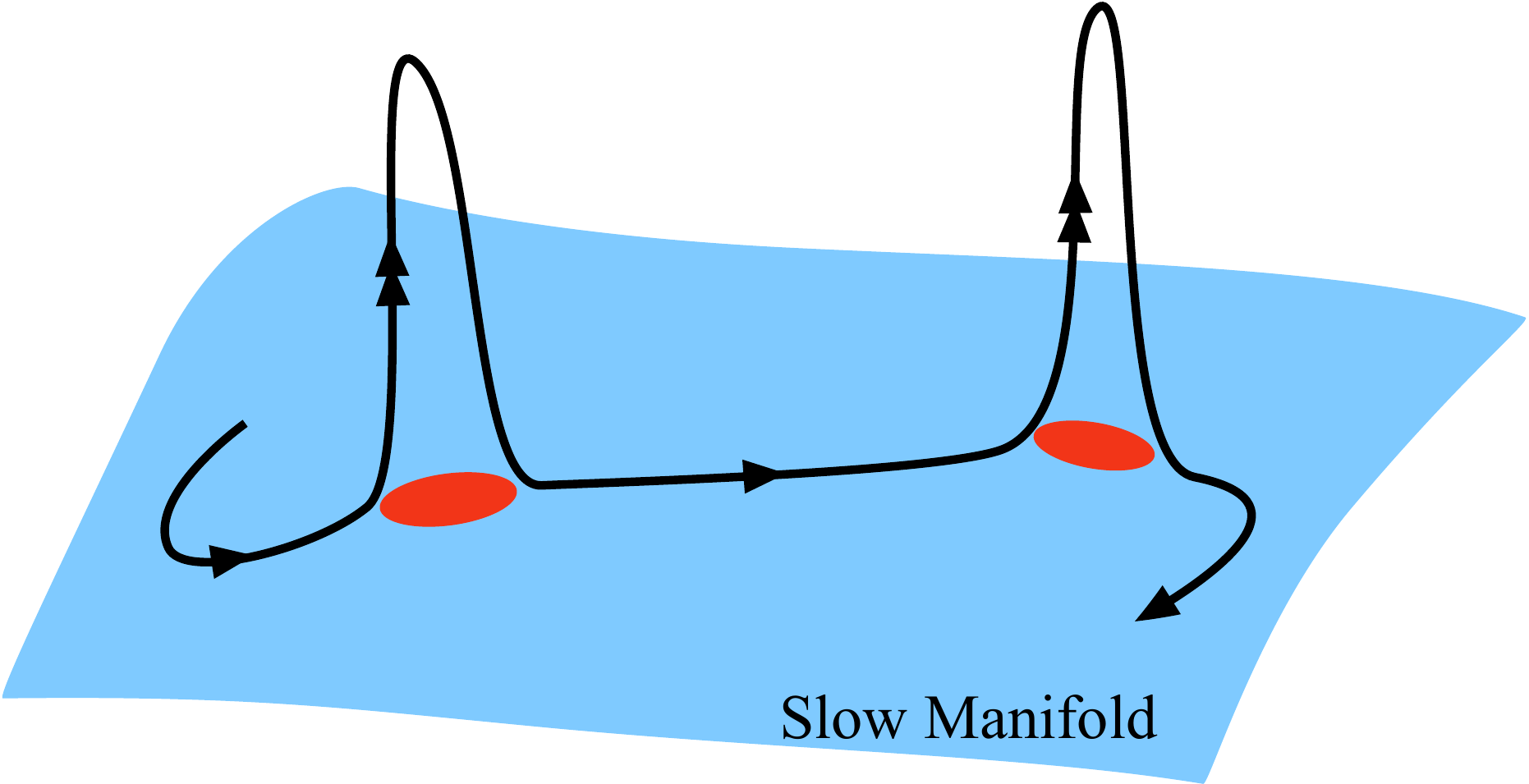}
\caption{A schematic picture of the 
a trajectory of a slow-fast system. The slow manifold is of the saddle type, that is, it consists of a
normally attracting component $\mathcal M_a$ (blue) and a normally repelling component $\mathcal M_r$ (red).
The trajectory diverges rapidly away from the slow manifold when it visits a repelling subset.
Subsequently the trajectory approaches the slow manifold along its attracting component.}
\label{fig:schem_slowfast}
\end{figure}

We first discuss the phenomenology of bursting in slow-fast systems and then 
demonstrate its implications on a concrete example. Figure~\ref{fig:schem_slowfast} sketches
the phase space geometry of a slow-fast system. It has an invariant slow manifold
where the trajectories follow the slow time scale. In the directions transverse to the
slow manifold, the dynamics follow the fast time scales. We assume that the 
slow manifold is normally hyperbolic. Loosely speaking, normal hyperbolicity means
that the transverse attraction to and repulsion from the manifold is stronger than
its internal dynamics~\cite{wiggins1994normally,jones95}. We also assume that the slow manifold is of the saddle type, that is,
it consists of two components: attracting and repelling.
Normal perturbations to the manifold on its attracting component decay 
over time while the perturbations over the repelling component grow.
Due to invariance of the slow manifold, trajectories starting on the manifold remain
on it for all times unless they exit the manifold through its boundaries. 

Now consider a trajectory that starts slightly off the slow manifold over its attracting 
component (the black curve in figure~\ref{fig:schem_slowfast}). Initially this trajectory
converges towards the slow manifold until it approaches its repelling component. At this point,
the normal repulsion pushes the trajectory away from the slow manifold where the fast time scales are
manifest. This rapid repulsion continues until the trajectory leaves the neighborhood of the
repelling component and is pulled back towards the attracting components. 

If the normal repulsion is strong enough, the episodes where the trajectory
travels away from the slow manifold appear as rapid bursts. The repelling subset of the
slow manifold can be a very complex set as a result of which the burst can appear chaotic
and sporadic. 

We demonstrate the bursting in slow-fast systems on a 
normal form of a singular Hopf bifurcation~\cite{guck12},
\begin{align}
\epsilon\, \dot x & = y-x^2-x^3\nonumber\\
\ \dot y &= z-x\nonumber\\
\dot z & = - \nu - a x - b y - c z, 
\label{eq:singHopf}
\end{align}
where $\epsilon\geq 0$ is a small parameter. For our discussion we fix the 
remaining parameters, $a=-0.3872$, $b=-0.3251$, $c=1.17$ and $\nu=0.0072168$.

\begin{figure}
\centering
\includegraphics[width=.45\textwidth]{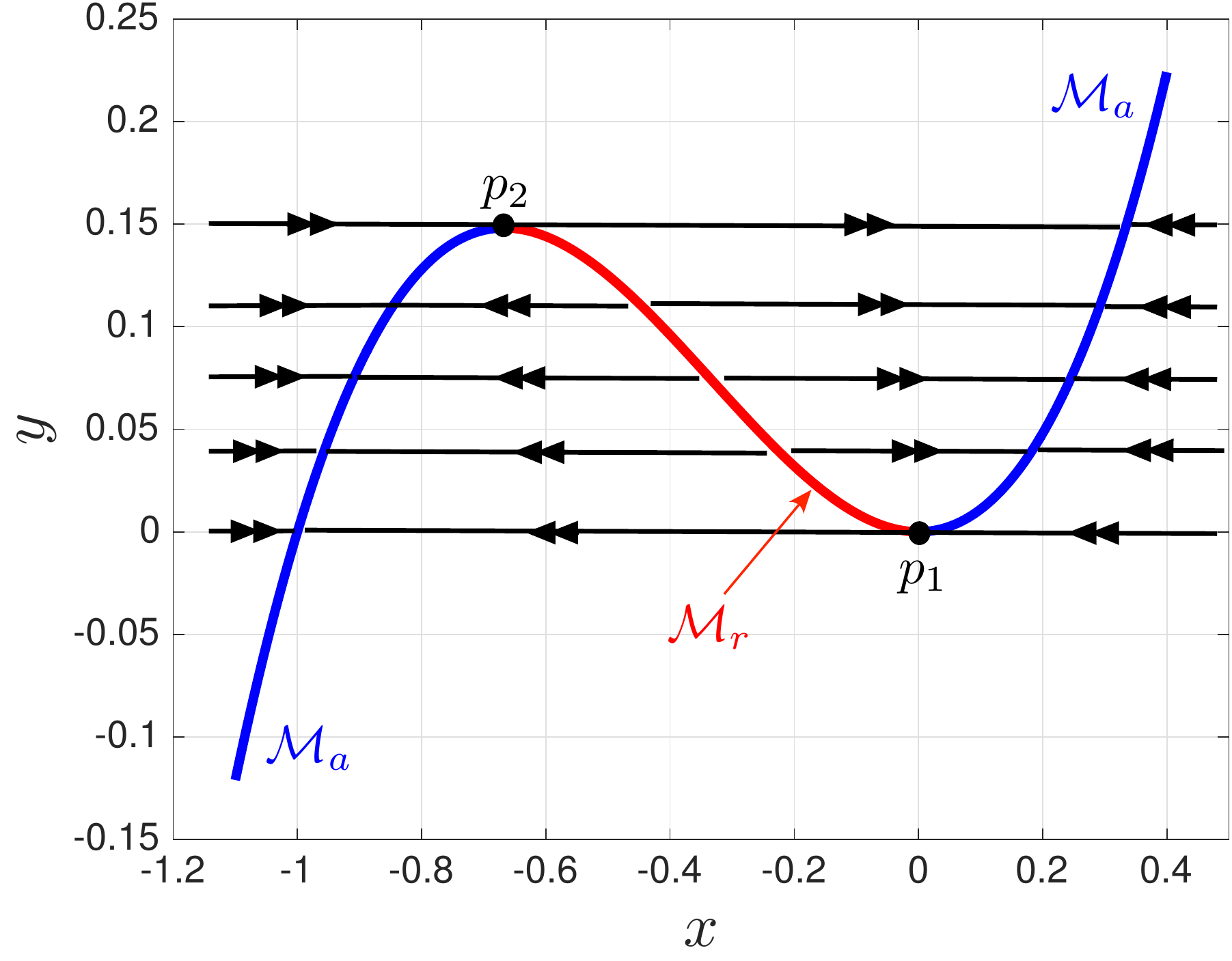}
\caption{Orbits of the rescaled system~\eqref{eq:singHopf_fast} with $\epsilon=0$
projected on the $x-y$ plane.
The s-shaped curve marks the slow manifold $y=x^2+x^3$ which consists of three
connected segments: two segments are attracting (blue) and one segment is repelling (red).}
\label{fig:mmo_sing_fast}
\end{figure}

We first discuss the singular limit where $\epsilon=0$.
In this limit, system~\eqref{eq:singHopf} reduces
to the differential-algebraic equations 
\begin{align}
0 & = y-x^2-x^3\nonumber\\
\dot y &= z-x\nonumber\\
\dot z & = - \nu - a x - b y - c z. 
\label{eq:singHopf_reduced}
\end{align}
This reduced system describes the slow flow on the \emph{critical manifold} 
\begin{equation}
\mathcal M_0 = \{(x,y,z): y=x^2+x^3\}.
\label{eq:critman}
\end{equation}
In order to discern the dynamics 
outside the critical manifold, we use a blow-up construction by 
rescaling time according to $t=\epsilon\tau$. 
The derivative with respect to the fast time $\tau$ is given
by $\frac{\id}{\id \tau} = \epsilon\frac{\id}{\id t}$. With this change of
variable, equation~\eqref{eq:singHopf} becomes
\begin{align}
x' & = y-x^2-x^3\nonumber\\
y' &= \epsilon(z-x)\nonumber\\
z' & = \epsilon(- \nu - a x - b y - c z), 
\label{eq:singHopf_fast}
\end{align}
where the prime denotes derivative with respect to the fast time $\tau$.
In the singular limit, $\epsilon=0$, we have $y'=0$ and $z'=0$.
Moreover, in the rescaled system, every point on the critical manifold is a fixed point
since $x'=y-x^2-x^3=0$. This is an artifact of the rescaling $t=\epsilon\tau$ 
which is singular at $\epsilon=0$. More precisely, points on the critical 
manifolds are fixed points with respect to the fast time scale. 
In turn, the slow dynamics on the critical manifold is 
given by the reduced system~\eqref{eq:singHopf_reduced}.
The combination of the reduced system~\eqref{eq:singHopf_reduced}
and the rescaled system~\eqref{eq:singHopf_fast} describes the motion on the critical
manifold and away from it.

\begin{figure*}
\centering
\includegraphics[width=.9\textwidth]{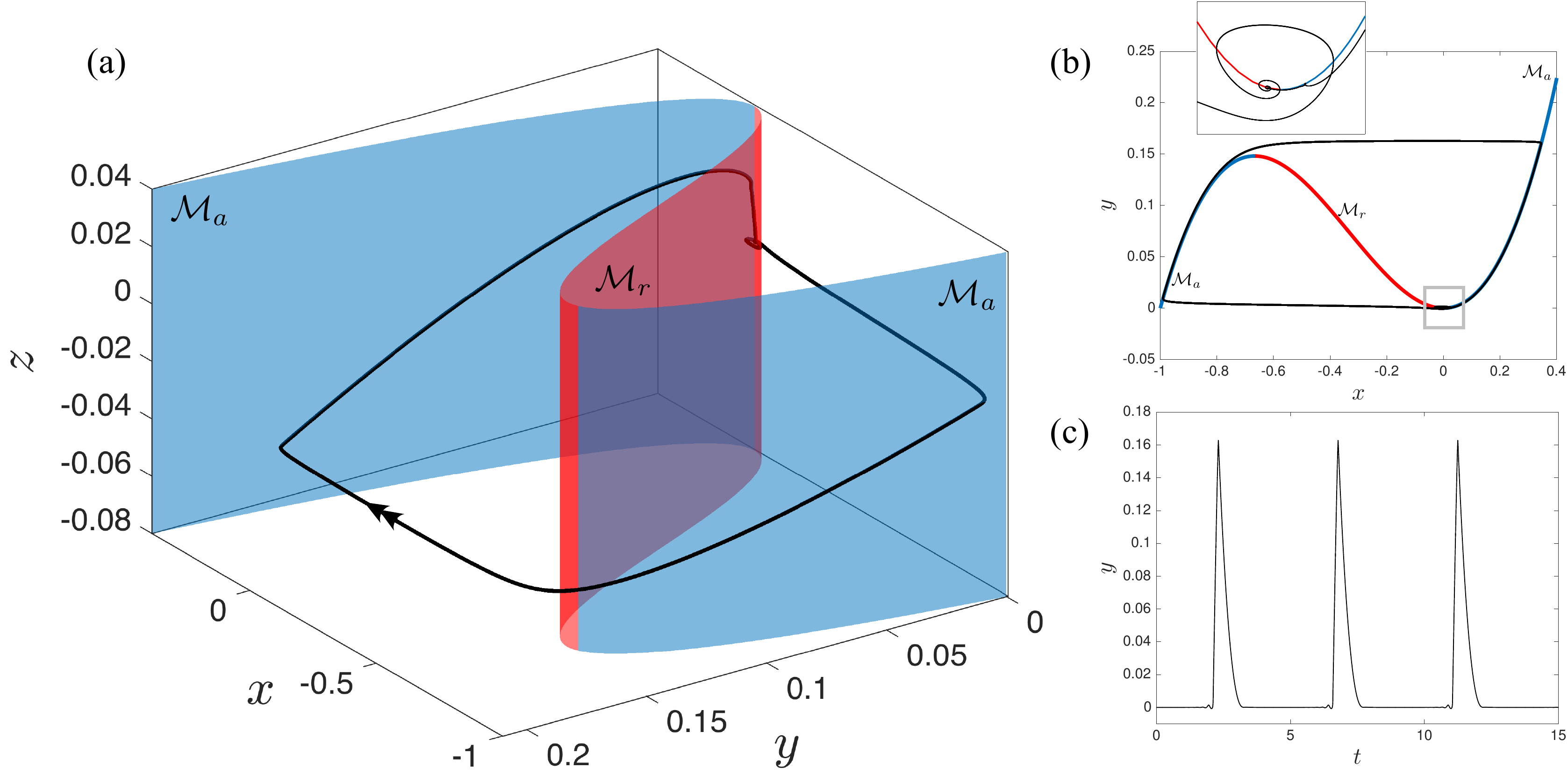}
\caption{The state space geometry of system~\eqref{eq:singHopf} with parameters
$(a,b,c,\nu,\epsilon)=(-0.3872,-0.3251,1.17,0.0072168,0.001)$.
(a) A stable periodic orbit of the system (black curve) is shown together with the critical manifold~\eqref{eq:critman}.
The attracting parts of the manifold are colored in blue and the repelling part is colored in red.
(b) Projection of panel (a) onto the $x-y$ plane. The inset shows a closeup view of the region enclosed in
a gray box.
(c) Time series of the $y$-coordinate along the periodic orbit.}
\label{fig:mmo}
\end{figure*}

Of particular relevance to us is the behavior of trajectories 
in a small neighborhood of the critical manifold. The critical manifold consists of 
three connected components (see figures~\ref{fig:mmo_sing_fast} and~\ref{fig:mmo}). 
Two of these components, denoted by 
$\mathcal M_a$, are normally 
attracting, meaning that trajectories starting away from them in a transverse direction converge
towards the critical manifold. In contrast, transverse perturbations to the normally repelling segment $\mathcal M_r$
diverge rapidly from the critical manifold. As a result, trajectories starting near the
repelling submanifold $\mathcal M_r$ are repelled to a neighborhood of the 
attracting manifold $\mathcal M_a$ where they follow the slow time scales along the 
critical manifold until they reach one of the fold points $p_1$ or $p_2$ (see figure~\ref{fig:mmo_sing_fast}).
At the folds, located on the boundary between the attracting and repelling submanifolds, 
the trajectory is repelled again from $\mathcal M_r$ toward the second segment of the 
attracting submanifold $\mathcal M_a$. This cycle continues indefinitely, creating
bursting trajectories that are repelled away from the repelling submanifold and
attracted back towards the critical manifold along its attracting segment. 

Now we turn our attention to the nonsingular case where $\epsilon>0$. 
The above analysis of the singular flow ($\epsilon=0$) bears some relevance to the 
nonsingular case ($\epsilon>0$). For sufficiently small perturbations, $0<\epsilon\ll 1$, 
the Geometric Singular Perturbation Theory (GSPT)~\cite{fenichel1979geometric}
guarantees, under certain conditions, that the critical manifold $\mathcal M_0$ survives as a perturbed invariant manifold
$\mathcal M_\epsilon$, that $\mathcal M_\epsilon$ is as smooth as the critical manifold,
and that $\mathcal M_\epsilon$ is $\mathcal O(\epsilon)$ close to the critical manifold $\mathcal M_0$. Furthermore, the normally
attracting or repelling properties of the perturbed manifold $\mathcal M_\epsilon$ is similar to that of the critical manifold
$\mathcal M_0$.

In particular, for system~\eqref{eq:singHopf}, the critical manifold $\mathcal M_0$
deforms into a nearby slow manifold $\mathcal M_\epsilon$. The perturbed slow manifold
has its own repelling and attracting submanifolds similar to those of $\mathcal M_0$ which
create bursting repulsion from and attraction towards the slow manifold. Figure~\ref{fig:mmo} shows a 
trajectory of the system for $\epsilon=10^{-3}$. At this parameter values, the system has undergone a 
supercritical Hopf bifurcation~\cite{guck08} giving birth to a stable periodic orbit (the black curve). This periodic orbit
carries much of the bursting properties described above for the singular system. Most of the time, the trajectory spirals 
outwards near the fold $p_1$. At some point, the trajectory approaches the repelling segment of the 
slow manifold $\mathcal M_\epsilon$ whereby it is repelled away towards its attracting segment.
The trajectory follows the attracting segment until it reaches the fold $p_2$ where the 
repelling segment again repels the trajectory towards the second attracting segment. Following
the attracting segment, the trajectories returns towards the fold $p_1$ and the small spiral motion repeats.
This cycle continues indefinitely. Figure~\ref{fig:mmo}(c) shows the time series of the 
$y$-coordinate along the periodic orbit showing the bursts resulting from repulsion away from the 
slow manifold.

For illustrative purposes, we presented in figure~\ref{fig:mmo} a parameter set 
where the asymptotic motion of the system is relatively simple, dictated by a single stable periodic orbit.
The dynamics is not always so predictable.
There are parameter values $(a,b,c,\nu,\epsilon)$ where the system undergoes 
period doubling bifurcations resulting in several co-existing unstable periodic orbits. 
As a result, a generic trajectory never settles down to a particular periodic orbit. 
Instead, it indefinitely bounces back and forth between unstable periodic orbits.
As a result, the bursting time series appear chaotic and less predictable.

We illustrate this on a system which exhibits chaotic bursts for a wide range of 
parameters.  Consider the coupled FitzHugh--Nagumo units~\cite{ansmann2013},
\begin{align}
\dot x_i = &  x_i(a_i-x_i)(x_i-1)-y_i+k\sum_{j=1}^{n} A_{ij}(x_j-x_i),\nonumber\\
\dot y_i = & b_ix_i-c_iy_i,
\label{eq:FHN}
\end{align}
where $n$ is the number of units and $(a_i,b_i,c_i)$ are constant parameters.
The units are coupled to each other through the summation term. The matrix $A$
with entries $A_{ij}\in\{0,1\}$ is the adjacency matrix that determines which units are coupled.
The constant $k$ determines the strength of the couplings. 
\begin{figure*}
\centering
\subfloat[]{\includegraphics[width=.33\textwidth]{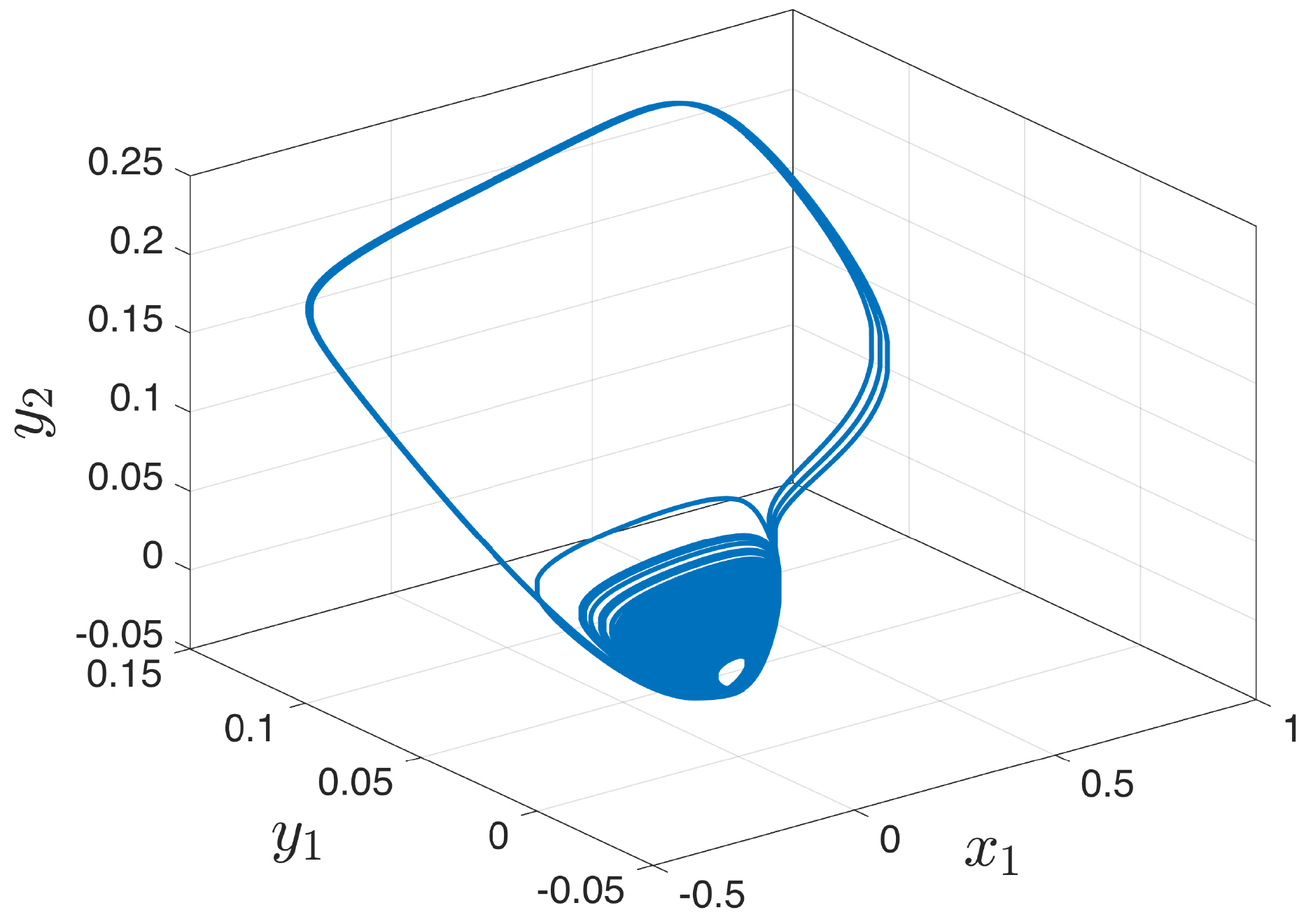}}
\subfloat[]{\includegraphics[width=.3\textwidth]{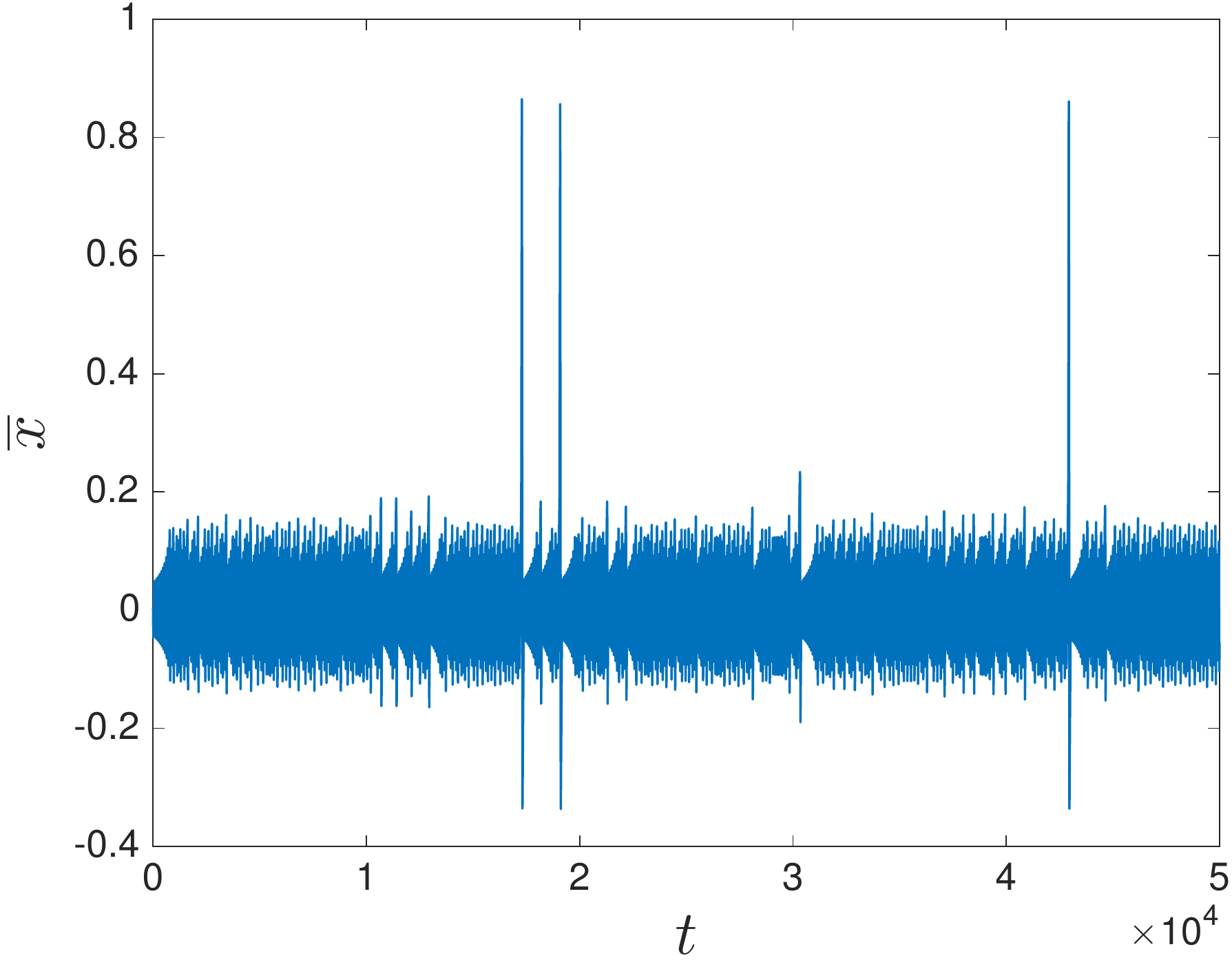}}
\subfloat[]{\includegraphics[width=.3\textwidth]{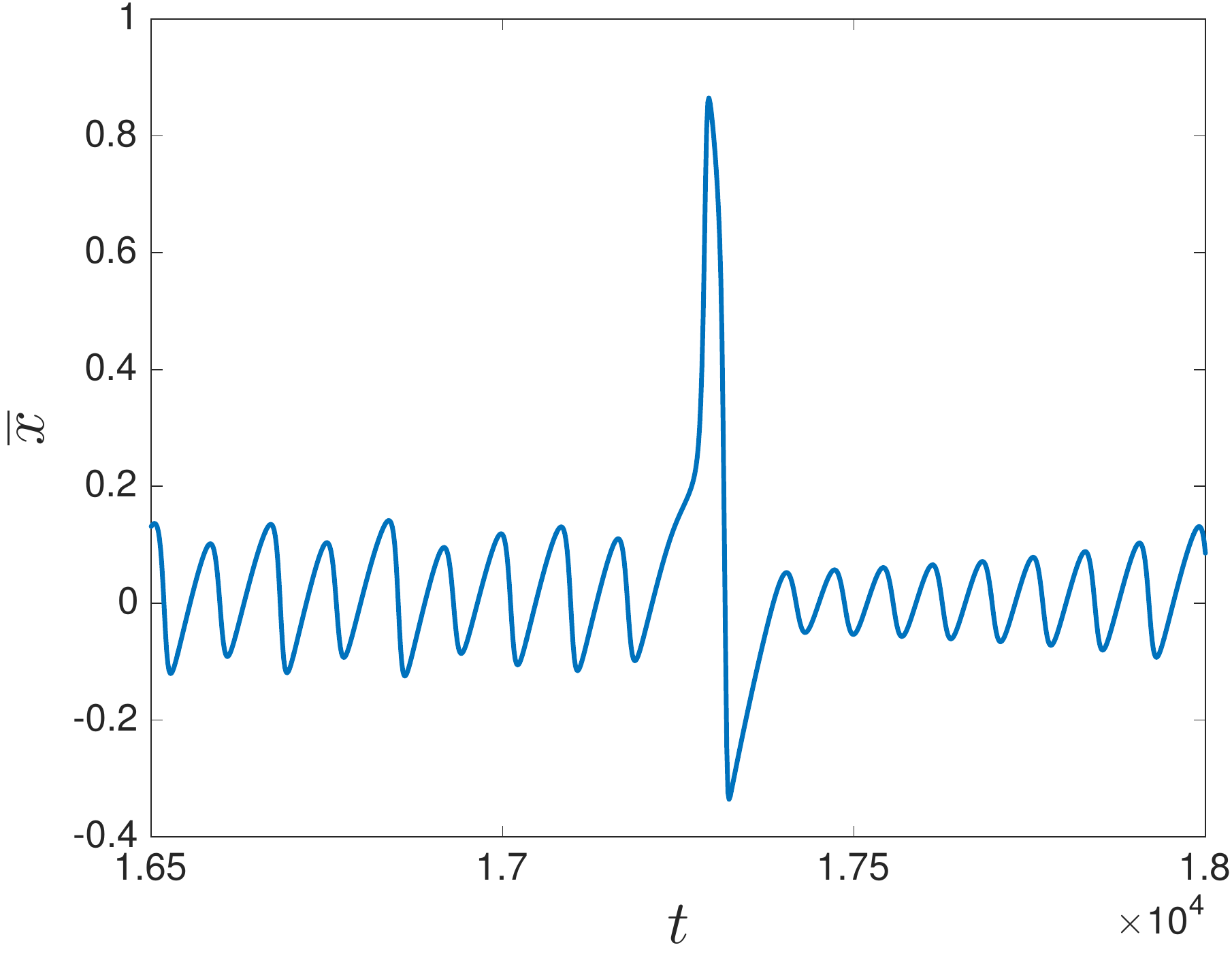}}
\caption{The FitzHugh--Nagumo oscillators~\eqref{eq:FHN} with two units ($n=2$). The parameters are
$a_1 =a_2 =-0.025794$, $c_1 = c_2 =0.02$, $b_1 = 0.0065$, $b_2 = 0.0135$ and $k=0.128$. 
The adjacency matrix $A$ is symmetric
with entries $A_{11}=A_{22}=0$ and $A_{12}=A_{21}=1$.
(a) A trajectory of the system projected onto the $(x_1,y_1,y_2)$ subspace.
(b) Time series of the observable $\overline x = \frac{1}{n}\sum_{i=1}^{n}x_i$.
(c) A closeup view of the first burst of $\overline{x}$.
}
\label{fig:FHN}
\end{figure*}

Figure~\ref{fig:FHN} shows a typical trajectory of the FitzHugh--Nagumo system
with two units ($n=2$). Also shown is the time series of the mean of $x_i$, i.e.
$\overline{x}=\frac{1}{n}\sum_{i=1}^{n}x_i$. The mean $\overline x$
mostly oscillates chaotically around $0$ with a relatively small variance. 
Once in a while, however, it exhibits relatively large excursions away from $0$ in the form of
bursts. As opposed to the periodic extreme events of figure~\ref{fig:mmo}, these bursts appear chaotically, with no regular recurrent pattern.
Similar extreme events have been observed in the FitzHugh--Nagumo system
with larger number of units and various parameter values~\cite{karnatak2014}.

In this chaotic regime, the geometry of the invariant sets and their
stable and unstable manifolds can be incredibly complex. 
One of the recent contributions to the field of slow-fast systems has been the development of
accurate numerical methods for computing such 
invariant manifolds~\cite{Guck04,Guck05,Lebiedz,Castelli2015}.
The computational cost of these manifolds increases with the dimension of the system such that
their computation is currently limited to four or five dimensional systems~\cite{babaee17}.
Nonetheless, understanding the mechanism behind extreme events in 
prototypical low-dimensional slow-fast systems has been 
greatly informative at the conceptual level.

\subsection{Homoclinic and heteroclinic bursting}\label{sec:clinic}
\begin{figure}
\centering
\subfloat[]{\includegraphics[width=.22\textwidth]{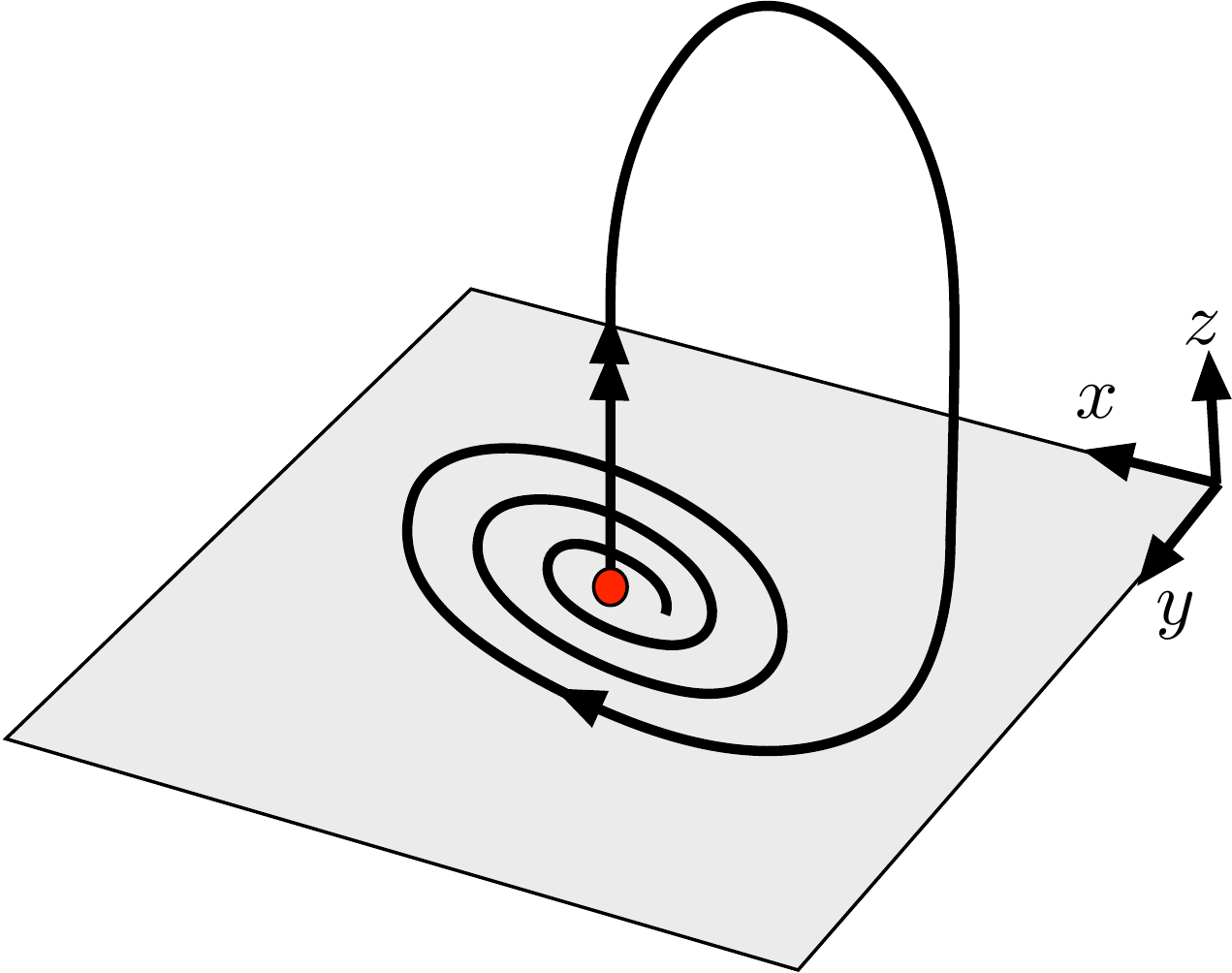}}
\subfloat[]{\includegraphics[width=.26\textwidth]{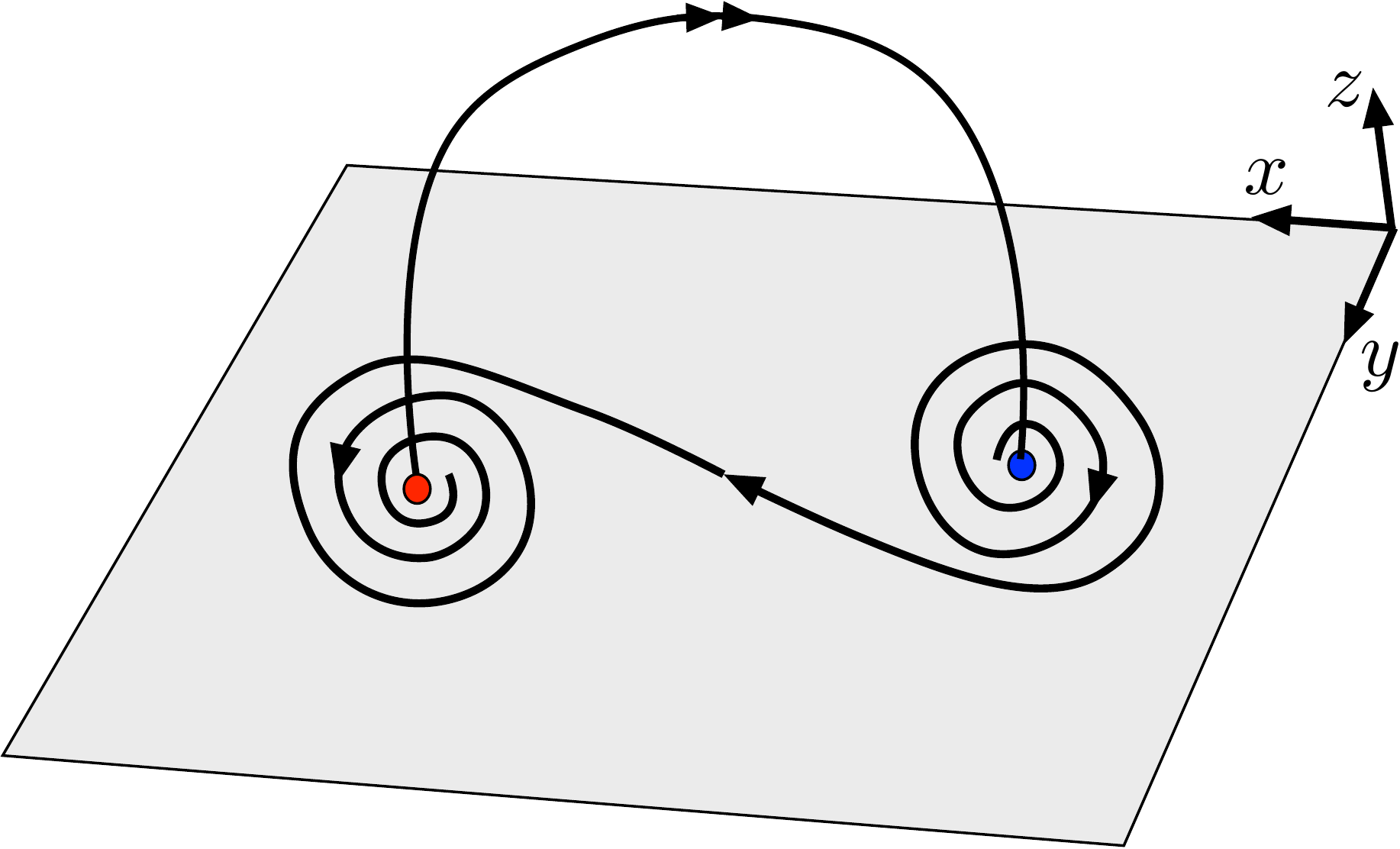}}
\caption{Sketches of a homoclinic (a) and a heteroclinic (b) orbit.}
\label{fig:homoHetero}
\end{figure}

Another geometric mechanism of generating extreme events is through homoclinic and heteroclinic connections
(see figure~\ref{fig:homoHetero}). Since these mechanisms share many of the characteristics of the slow-fast systems
discussed in section~\ref{sec:slowfast}, we limit this section to a brief discussion of the main ideas underlying homoclinic and
heteroclinic bursting.

An example of a homoclinic connection is that of the Shilnikov orbit of a saddle-focus fixed point. 
This is an unstable fixed point with a two-dimensional 
stable manifold and a one-dimensional unstable manifold (see figure~\ref{fig:homoHetero}(a)). Within the stable manifold, the trajectories
spiral towards the fixed point while they are repelled from the fixed point in its unstable direction. 
The Shilnikov orbit is the homoclinic trajectory that is asymptotic to the fixed point both
in forward time and in backward time. 

Shilnikov~\cite{shilnikov1965} proved that, if the attraction within the stable manifold is weaker than the repulsion along the unstable manifold,
small perturbations to the system give birth to infinitely many unstable periodic orbits around the 
homoclinic orbit. These periodic orbits resemble the shape of the original Shilnikov orbit~\cite{gaspard1983}. More precisely,
the periodic orbits consist partly of spiral motion towards the fixed point and partly of bursting motion away from it. 
Generic trajectories shadow these periodic orbits such that the time series of their $z$-component
exhibits small scale oscillations, corresponding to the spiraling motion, and occasional bursts, corresponding to 
repulsion along the unstable manifold~\cite{cvitanovic1989,cvitanovic1991}. 
Since the periodic orbits are all unstable, the motion along generic trajectories can be 
chaotic resulting in very complex dynamics. 
A classical example of such chaotic motion is the Rossler attractor~\cite{rossler1976,letellier1995}. 

Although the Shilnikov bifurcation was first studied as a route to chaotic motion in simple systems, 
it has found many applications in explaining the self-sustained bursting phenomena 
observed in nature. These include, for instance, sudden variations in geophysical flow patterns~\cite{meacham2000,timmermann2003},
spiking and synchronization in neural networks~\cite{ermentrout1998,izhikevich2000,izhikevich2003} and chemical reactions~\cite{elezgaray1992}.

A similar mechanism of bursting is through heteroclinic connections. As opposed to the homoclinic case, the heteroclinic orbit asymptotes to different fixed points in forward and backward times.
Figure~\ref{fig:homoHetero}(b), for instance, depicts a heteroclinic connection corresponding to the phase space of a three-dimensional vector field introduced in Ref.~\cite{PRE2016}.
As in the homoclinic case, the heteroclinic bursting has been useful in explaining several spiking behavior observed in nature
from nonlinear waves to turbulent fluid flow~\cite{coller1994,Jones1994,han1995,haller1995,coller1997,kawahara2001,faraz_adjoint}.

\subsection{Noise-induced transitions}\label{sec:tpt}
So far we have discussed deterministic systems
which possess a self-sustaining mechanism for generating extreme events. 
However, an important class of rare extreme events are induced by noise~\cite{horsthemke1984,Broeck1994,neiman2002,moore1}; see also \cite{moore0} for an excellent review for this form of transitions. Such systems
typically have equilibria that are stable in the absence of noise. Noise, however,
makes it possible to transition from the neighborhood of one equilibrium to the other. 

In such systems the transition mechanism is the noise and, as such, there is
no ambiguity regarding what underlies the rare events. However, the route the
system takes during each transition is not as clear. In fact, due to the random nature
of the system, the transition routes can only be identified probabilistically. 
In particular, one can inquire about the most likely route the system takes in 
traveling between two states. The answer facilitates the prediction of individual transitions
as well as the quantification of transition rates in an ensemble of experiments. 
In this section, we briefly review the \emph{transition-path theory} which is a framework for addressing these questions.

The origins of the transition-path theory
stem from chemical physics where one is interested in computing the rate of
chemical reactions that lead to a transition from the reactant state to 
the product state~\cite{wigner1938,horiuti1938,eyring1935,yamamoto1960,chandler1978,pratt1986}.

The transition-path theory aims to go beyond computing the transition rates and determines the
most likely paths that the system may take during the transitions. To describe this theory, 
we consider the Langevin equation,
\begin{equation}
m \ddot x = -\nabla V(x) -\gamma \dot x + \sqrt{2}\sigma(x)\eta(t),
\label{eq:Langevin_01}
\end{equation}
where $u=(x,\dot x)\in \mathbb R^{2n}$ determines the state of the system,
$m$ is the mass matrix, $V:\mathbb R^n\to\mathbb R$ is the potential and $\gamma$ is 
the friction coefficient. The stochastic process $\eta(t)$ is a white noise with mean zero and
covariance $\langle \eta_i(t)\eta_j(s)\rangle=\delta_{ij}\delta(t-s)$ and
$a(x)=\sigma(x)\sigma(x)^\dagger\in\mathbb R^{n\times n}$ is the diffusion matrix. 

For our introductory purposes it is helpful to consider the over-damped case $\gamma\gg 1$
where equation~\eqref{eq:Langevin_01} reduces to 
\begin{equation}
\dot u = -\nabla V(u)+\sqrt 2\sigma (u) \eta(t),
\label{eq:Langevin_02}
\end{equation}
where $u=x\in\mathbb R^n$ determines the state of the system. For simplicity, we have assumed $m=\mbox{Id}$
and rescaled time to eliminate the dependence on the friction coefficient $\gamma$.

Figure~\ref{fig:transition_geom} sketches the over-damped system~\eqref{eq:Langevin_02}
in the one-dimensional case ($n=1$). In absence of noise $\eta(t)$, the system converges asymptotically to one of the
two local minima of the potential $V$. The noise, however, nudges the trajectory away from these equilibria. 
In rare instances, the trajectory can even pass over the saddle separating the two equilibria
causing a transition from equilibrium $u_1$ towards equilibrium $u_2$ and vice versa. 

\begin{figure}
\centering
\includegraphics[width=.45\textwidth]{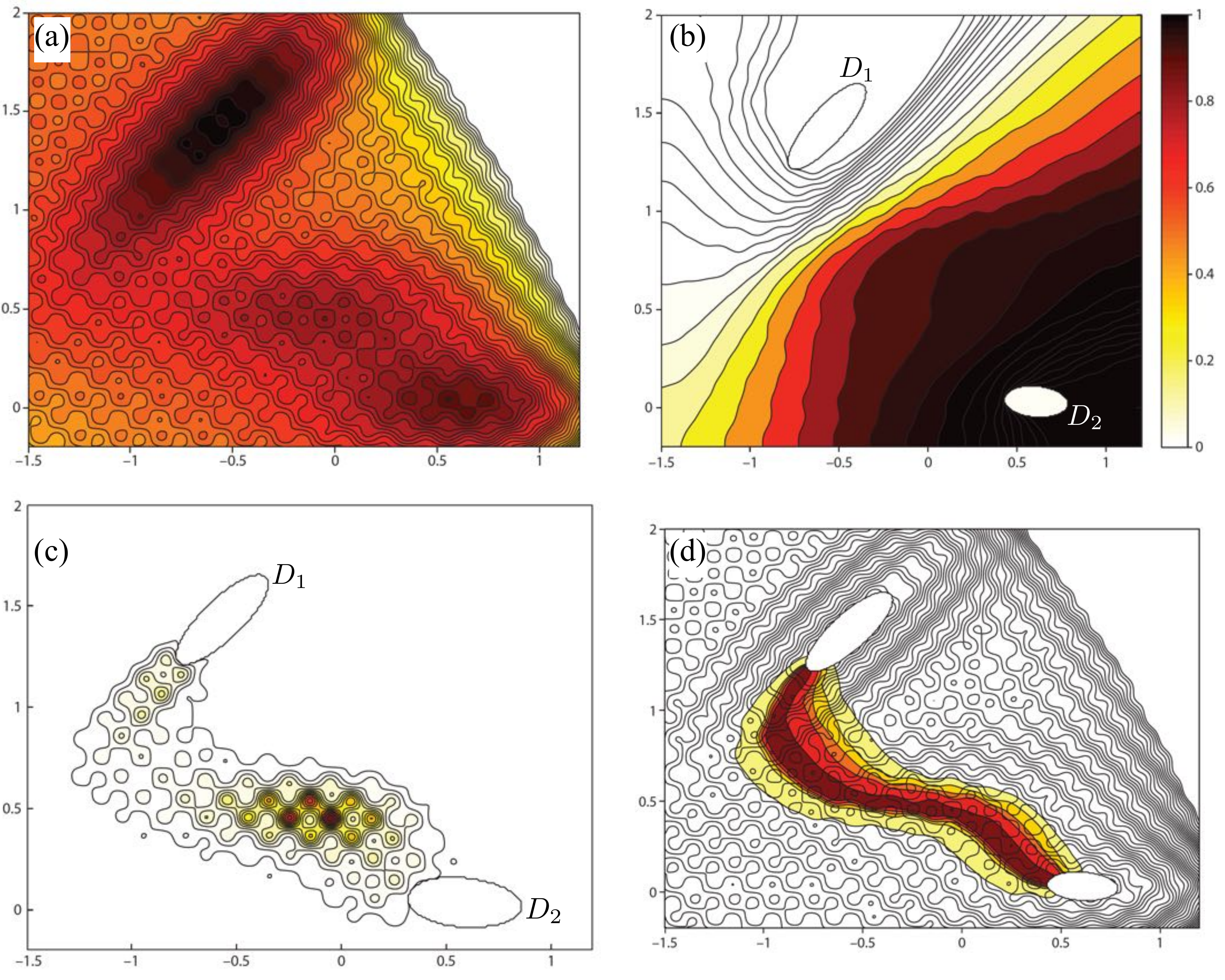}
\caption{Two-dimensional rugged Mueller potential. 
(a) Contour lines of the potential $V$. Darker colors mark smaller values. 
(b) Contour lines of a committor function corresponding to the rugged Muller potential.
(c) Contour lines of the corresponding probability density $\rho_{12}$ of transition trajectories.
(d) The flow lines of the probability current $J_{12}$ of the transition trajectories.
Figure reproduced from Ref.~\cite{weinan2010}}
\label{fig:rm_potential}
\end{figure}

In this simple example, there is no ambiguity about the path these rare transitions take since
there is only one degree of freedom available. However, in higher dimensional problems ($n\geq 2$),
the rare transitions have more options for traveling between local minima of the potential $V$. 
Figure~\ref{fig:rm_potential}(a), for instance, shows the so-called rugged Mueller potential in two dimensions ($n=2$)
with infinitely many possible paths between any pair of local minima.

In this case, the following question arises: Given two sets $D_1,D_2\subset\funcspace$, what is the most likely path 
the system may take for transitioning from set $D_1$ to set $D_2$? Figure~\ref{fig:rm_potential}(b)
shows two such sets that cover the the lowest valleys of the potential $V$.
To answer this question, we assume that system~\eqref{eq:Langevin_02} is ergodic with
a unique invariant probability density $\rho:\mathcal U\to\mathbb R^+$ such that
the probability density of observing the state $u$ is $\rho(u)$. We would like to find the probability 
density $\rho_{12}(u)$ which corresponds to the probability that a trajectory passing through $u$
has come from $D_1$ and will be going to $D_2$. This probability density is given by
\begin{equation}
\rho_{12}(u)=q_+(u)q_-(u)\rho(u),
\label{eq:rho12}
\end{equation}
where $q_-,q_+:\funcspace\to\mathbb [0,1]$ are the so-called committor functions. 
The committor function $q_-(u)$ measures the probability that 
a trajectory passing through $u$ came from $D_1$. On the other hand, $q_+(u)$
measures the probability that the trajectory passing through $u$ will arrive at $D_2$ before
arriving at $D_1$.

One can show that the committor functions satisfy the backward Kolmogorov equations
\begin{equation}
\begin{cases}
Lq_+ = 0,& u\in \funcspace\backslash (D_1\cup D_2),\\
q_+ = 0,& u\in D_1,\\
q_+ = 1,& u\in D_2,\\
\end{cases}
\label{eq:q+}
\end{equation}
and 
\begin{equation}
\begin{cases}
L^\dagger q_- = 0,& u\in \funcspace\backslash (D_1\cup D_2),\\
q_- = 1,& u\in D_1,\\
q_- = 0,& u\in D_2,\\
\end{cases}
\label{eq:q-}
\end{equation}
where $L=-\nabla V(u)\cdot \nabla +a(u):\nabla\nabla$ is a linear operator and $L^\dagger$ is
its adjoint with respect to the inner product $\langle \alpha,\beta\rangle =\int_{\funcspace} \alpha(u)\beta(u)\rho(u)\id u$ 
(see, e.g., Refs.~\cite{weinan2006,vanden2006}).
In addition, the density $\rho$ satisfies the forward Kolmogorov (or Fokker--Planck) equation,
\begin{equation}
\nabla\cdot (\rho\nabla V)+\nabla \nabla : (\rho a)=0,
\label{eq:rho}
\end{equation}
where the time derivative vanishes since the density is invariant. 

In order to evaluate the transition probability density $\rho_{12}$, 
one needs to solve equations~\eqref{eq:q+},~\eqref{eq:q-} and~\eqref{eq:rho}
for $q_+$, $q_-$ and $\rho$, respectively. Then the transition probability density is
computed from~\eqref{eq:rho12}. Figure~\ref{fig:rm_potential}(c) shows the
transition probability density $\rho_{12}$ corresponding to the rugged Mueller potential. 

Recall that the probability density $\rho_{12}(u)$ corresponds to the 
probability that a trajectory passing through $u$
has come from $D_1$ and will be going to $D_2$. Although useful, 
this probability density is still a pointwise quantity which does not immediately 
inform us about the most likely \emph{path} the system will take in going from $D_1$ to
$D_2$.

To address this shortcoming, the transition-path theory uses the probability current 
$J_{12}:\funcspace \backslash (D_1\cup D_2)\to \mathbb R^n$
associated with the transition probability density $\rho_{12}$. The vector field $J_{12}$ is defined such that for any 
codimension-one surface $\mathcal S\in \funcspace \backslash (D_1\cup D_2)$ the integral of $J_{12}$
over the surface, i.e. $\int_{\mathcal S} J_{12}(u)\cdot \id S(u)$, equals the probability flux of 
transition trajectory through $\mathcal S$. The current $J_{12}$ can be expressed 
explicitly in terms of the quantities introduced previously~\cite{weinan2006,vanden2006} as
\begin{equation}
J_{12}=q_+q_-J +\rho q_-a\nabla q_+-\rho q_+a\nabla q_-,
\end{equation}
where $J=-\rho\nabla V-\nabla\cdot (\rho a)$ is the probability current associated with the probability density $\rho$.

Figure~\ref{fig:rm_potential}(d) shows the streamlines of the transition current $J_{12}$. The color
encodes the probability of the transition along each path such that the darker colors mark a higher transition probability. 
This figure finally shows the most probable path the transitions trajectories take in going from $D_1$ to $D_2$. 

Therefore, for noise-driven rare transitions, the transition-path theory provides a rigorous framework for 
computing the most likely mechanism for the rare events. We recall, however, that computing the transition paths in this
framework requires the solutions to three PDEs~\eqref{eq:q+},~\eqref{eq:q-} and~\eqref{eq:rho}. Solving these 
equations in higher dimensions are quite costly such that the applications of transition-path theory have been 
limited to two- and three-dimensional systems~\cite{Metzner2006}. We finally point out that a number of numerical methods 
for approximating the rare transition paths have been developed in order to partially remedy this 
high computational cost~\cite{gonzalez1989,bolhuis2002, dellago2002, weinan2002,maragliano2006,weinan2007,pan2008}.

\section{Variational method for physics-based probing of extreme events}\label{sec:probe}
In this section, we review a recent variational method 
for discovering the mechanisms that cause the extreme events. 
This method exploits the physics given by the governing equations~\eqref{eq:masterEq}
together with the statistical information from the system attractor in order to
find initial states $u_0$ that over a prescribed time interval 
develop into an extreme event. The hope is to learn about the 
mechanism that causes the extremes by examining the states that 
precede the extreme events. We first introduce the variational method 
in a general framework and then present two specific applications of the method.

\subsection{The variational method}
Consider an observable $\obs:\funcspace\to\mathbb R$ 
whose time series along the system~\eqref{eq:masterEq}
is known to exhibit extreme events (see Definition~\ref{def:ee}).
Also assume that there is a typical timescale $\tau$
over which the observable grows from its typical values and increases
past its extreme value threshold $\obs_e$. We therefore seek initial states $u_0\in\funcspace$
such that $\obs(\solmap{\tau}(u_0))>\obs_e$. This motivates the
definition of the \emph{domain of attraction of extreme events} as follows.

\begin{defn}[Extreme Event Domain of Attraction]
For an extreme event set $E_\obs(\obs_e)$ and a prescribed time $\tau>0$, the corresponding
finite-time domain of attraction to the extreme events 
is the set
\begin{align}
A_\obs(\tau,\obs_e)& =\{u\in\funcspace\backslash E_{\obs}(\obs_e): \exists\, t\in (0,\tau],\ \solmap{t}(u)\in E_\obs(\obs_e)\} \nonumber\\
&=\left[\bigcup_{0<t\leq \tau}\solmap{-t}(E_\obs(\obs_e))\right]\backslash E_\obs(\obs_e).
\end{align} 
\end{defn}

Here $\solmap{-t}(B)$ is shorthand for the pre-image $(\solmap{t})^{-1}(B)$
of a set $B\in\mathcal B$. The set $A_f(\tau,f_e)$ contains the states
$u$ that at some future time $t$, with $t\leq \tau$, enter the extreme event set $E_f(f_e)$.
We remove the extreme event set $E_f(f_e)$ from the domain of attraction to exclude the states that are extreme
at the initial time.

The extreme event domain of attraction $A_\obs$ can be an extremely complex set
whose numerical estimation is a daunting task. In addition, determination of the entire set
may be unnecessary for deciphering the mechanisms that give rise to extreme events. 
Instead, one representative state from this set may suffice in discovering the extreme event generating mechanism.

We proposed in~\cite{Farazmande1701533} to obtain the desired representative states as the solutions of a 
constrained optimization problem. In this approach, we seek states $u_0\in\funcspace$ that maximize 
the growth of the observable $\obs$ over a prescribed time interval of length $\tau>0$. 
More precisely, we seek the solutions to the maximization problem
\begin{equation}
\sup_{u_0\in\attr}\left[ \obs(\solmap{\tau}(u_0))-\obs(u_0)\right],
\label{eq:opt_general}
\end{equation}
where $\attr$ is a subset of $\funcspace$ to be discussed shortly.
There are two constraints that are embedded in the optimization problem~\eqref{eq:opt_general}.
One constraint is enforced through $\solmap{t}$ generated by the governing equations~\eqref{eq:masterEq}.
In other words, it is implicitly implied that $u(t)=\solmap{t}(u_0)$ is a solution of the governing equations. 

A second constraint is implied by requiring the state $u_0$ to belong to the subset $\attr$. 
We envision $\attr$ to approximate the attractor of the system~\eqref{eq:masterEq}.
\textit{This constraint is essential for discarding exotic states} that belong to the state space 
$\funcspace$ but have negligible probability of being observed under the natural dynamics 
generated by the governing equations. It is known that dissipative differential equations
often posses an attractor which is a subset of the state space~\cite{ruelle1989,constantin2012}. While the system
can be initialized from any arbitrary states $u_0\in\funcspace$, its trajectories quickly converges to the
attractor and remain on it. As a result, much of the function space $\funcspace$ is 
unexplored; the only states relevant to long term dynamics of the system are the 
ones belonging to the attractor or a small neighborhood of it. To this end, this additional constraint not only leads to more relevant states as precursors, but \textit{it also  reduces the computational cost
of the optimization problem}, since we explore only the physically relevant solutions.
For instance, the state space of the
FitzHugh--Nagumo system shown in figure~\ref{fig:FHN} is $\mathbb R^4$. However, it is visually appreciable that its trajectories 
converge to a small subset of $\mathbb R^4$.

Constraining the optimal states $u_0$ in equation~\eqref{eq:opt_general} to belong to
the attractor $\attr$ eliminates the states that may lead to a large growth of the observable
but are dynamically irrelevant. The FitzHugh--Nagumo system, for instance, has transient trajectories 
along which $\overline x$ becomes larger than $1.5$ which is much larger than the typical bursts shown 
in figure~\ref{fig:FHN}. These unusually large bursts, however, occur along trajectories that are away from the
attractor and therefore are not sustained. 

The attractor can be a very complex set whose estimation is quite difficult.
In fact, numerical approximation of the attractors even in low-dimensional systems is an
active area of research (see, e.g., Refs.~\cite{Dellnitz2004,chen12}). For our purposes an approximate representation of 
the attractor is sufficient. Here, we assume that the attractor can be approximated by the set
\begin{equation}
\attr = \{u_0\in\funcspace : \underline c_i\leq C_i(u_0)\leq \overline{c}_i,\quad i=1,2,\cdots, k \},
\end{equation}
where $k\in\mathbb N$ determines the number of constraints, the maps $C_i:\funcspace\to \mathbb R$
are smooth enough and $\underline{c}_i,\overline c_i\in \mathbb R$ are the lower and upper bounds of $C_i$.
The choice of the maps $C_i$ and their bounds depends on the problem and is elaborated in the following sections. 

With the two constraints discussed above, the optimization problem~\eqref{eq:opt_general}
can be written more explicitly as 
\begin{subequations}
\label{eq:optGen2}
\begin{equation}
\sup_{u_0\in\funcspace}\left[ \obs(u(\tau))-\obs(u_0)\right],
\end{equation}
\begin{equation}
\partial_t u = N(u),\quad u(0)=u_0,
\end{equation}
\begin{equation}
\underline c_i\leq C_i(u_0)\leq \overline{c}_i,\quad i=1,2,\cdots, k,
\end{equation}
\end{subequations}
where $u(t)$ is the shorthand notation for a trajectory of the system~\eqref{eq:masterEq}.
If the set $\attr$ is compact in $\funcspace$ and the observable $\obs$ and the solution map $\solmap{t}$
are smooth enough, then there exist solutions to problem~\eqref{eq:optGen2}.
These solutions are not necessarily unique. In fact, often there are multiple
local maxima which may or may not be informative as to the origins of the
extreme events. The relevance of the local minimizers can only be determined a posteriori.
There are standard numerical methods for approximating the solutions of the 
constrained optimization problems of the form~\eqref{eq:optGen2} that we do not
review here but refer the interested reader to Refs.~\cite{hinze2008,Herzog2010,faraz_cont,bertsekas2014}.

\begin{figure}
\centering
\includegraphics[width=.45\textwidth]{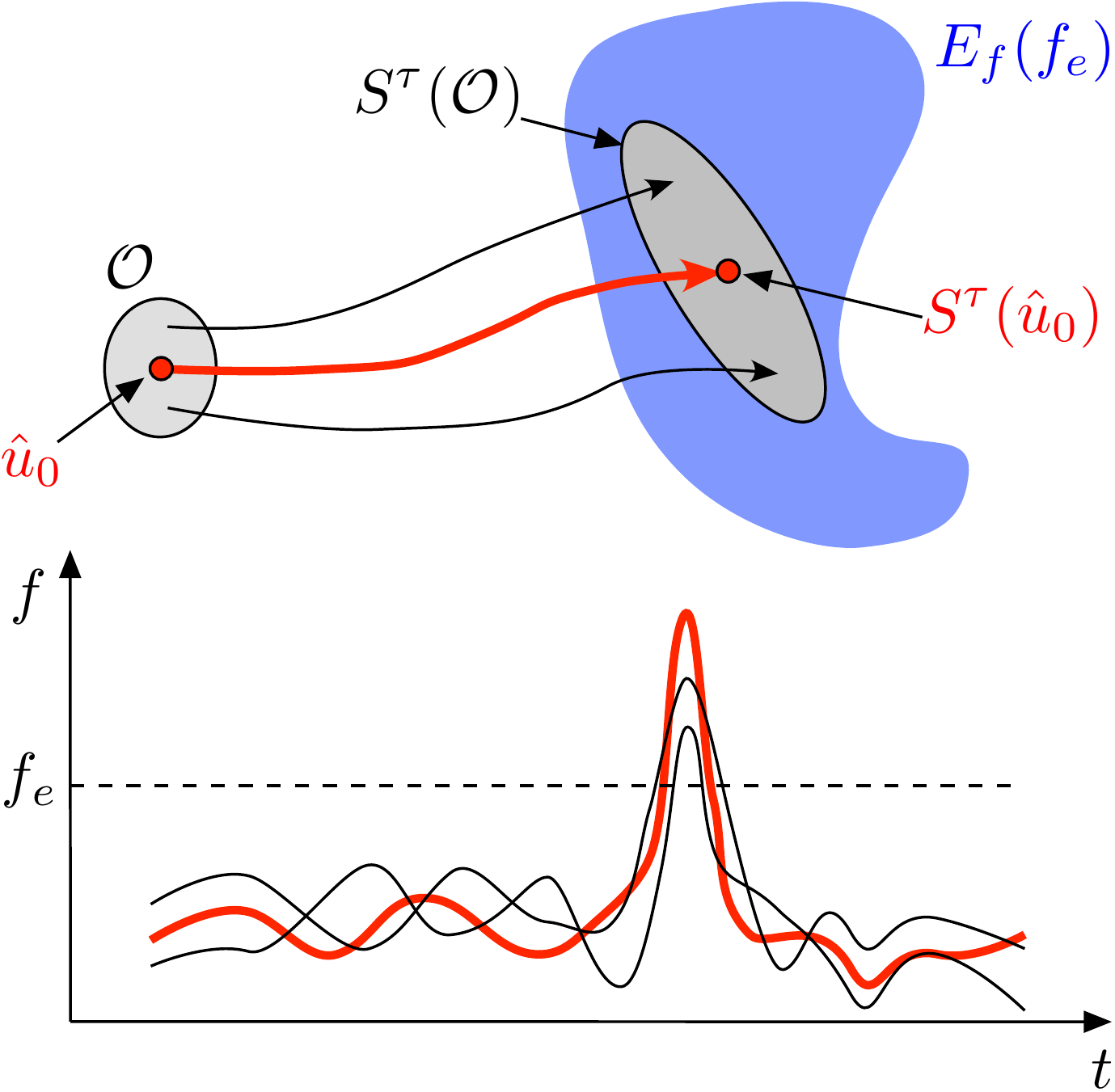}
\caption{Nearby trajectories to the optimal solution also give rise to extreme events.
The upper panel shows a solution $\hat u_0$ of the optimization problem~\eqref{eq:optGen2}
and the ensuing trajectory (red curve). Trajectories passing through a sufficiently small open neighborhood
$\mathcal O$ of $\hat u_0$ also give rise to extreme events. The lower 
panel depicts the evolution of the observable $\obs$ along these trajectories.}
\label{fig:continuity}
\end{figure}

Let $\hat u_0$ denote a solution of the problem~\eqref{eq:optGen2} corresponding
to an extreme event, i.e., $\obs(\solmap{\tau}(\hat u_0))>f_e$.
We point out that a generic trajectory of the system~\eqref{eq:masterEq}
may never exactly pass through the state $\hat u_0$. However, if the solution map
$\solmap{t}$ is continuous, any trajectory passing through a sufficiently small 
neighborhood of $\hat u_0$ will also develop into an extreme event. This is
illustrated in figure~\ref{fig:continuity}.

We demonstrate the application of this variational method on two examples. 
The first example involves the discovery of internal energy transfers that lead to 
the extreme energy dissipation episodes in a turbulent fluid flow. The second examples
involves prediction of unusually large ocean surface waves, commonly known as rogue waves. 

\subsection{Application to a turbulent fluid flow}\label{sec:kolm}
In this section, we present the application of the variational method to the
extreme energy dissipation in a turbulent fluid flow. This flow is analyzed in detail
in Ref.~\cite{Farazmande1701533}; here, we reiterate our main findings and 
add a number of complementary comments.
Consider the solutions to the two-dimensional incompressible Navier--Stokes equation
\begin{equation}
\partial_t u = -u\cdot \nabla u-\nabla p +\nu \Delta u + F,\quad \nabla\cdot u =0,
\label{eq:nse}
\end{equation}
where $u:\mathbb T^2\times \mathbb R^+\to \mathbb R^2$ is the velocity filed, $p:\mathbb T^2\to \mathbb R$
is the pressure field, $\nu$ is the kinematic viscosity and the torus $\mathbb T^2=[0,2\pi]\times [0,2\pi]$ is 
the fluid domain with periodic boundary conditions. The velocity $u(x,t)$ and pressure $p(x,t)$ are 
functions of the spatial variables $x=(x_1,x_2)\in\mathbb T^2$ and time $t\in\mathbb R^+$.
The flow is driven by the deterministic Kolmogorov forcing 
$F=\sin(k_fx_2)e_1$ where $k_f=4$ is the forcing wavenumber and $e_1=(1\quad 0)^\top$.
The simulations start from a random initial condition $u(x,0)$ which is in turn 
propagated forward in time by numerically integrating the Navier--Stokes equation~\eqref{eq:nse}.
We allow enough time elapse before collecting data in order to ensure
that the initial transients have decayed and the trajectory has settled to the system attractor.
\begin{figure*}
\centering
\subfloat[]{\includegraphics[width=.4\textwidth]{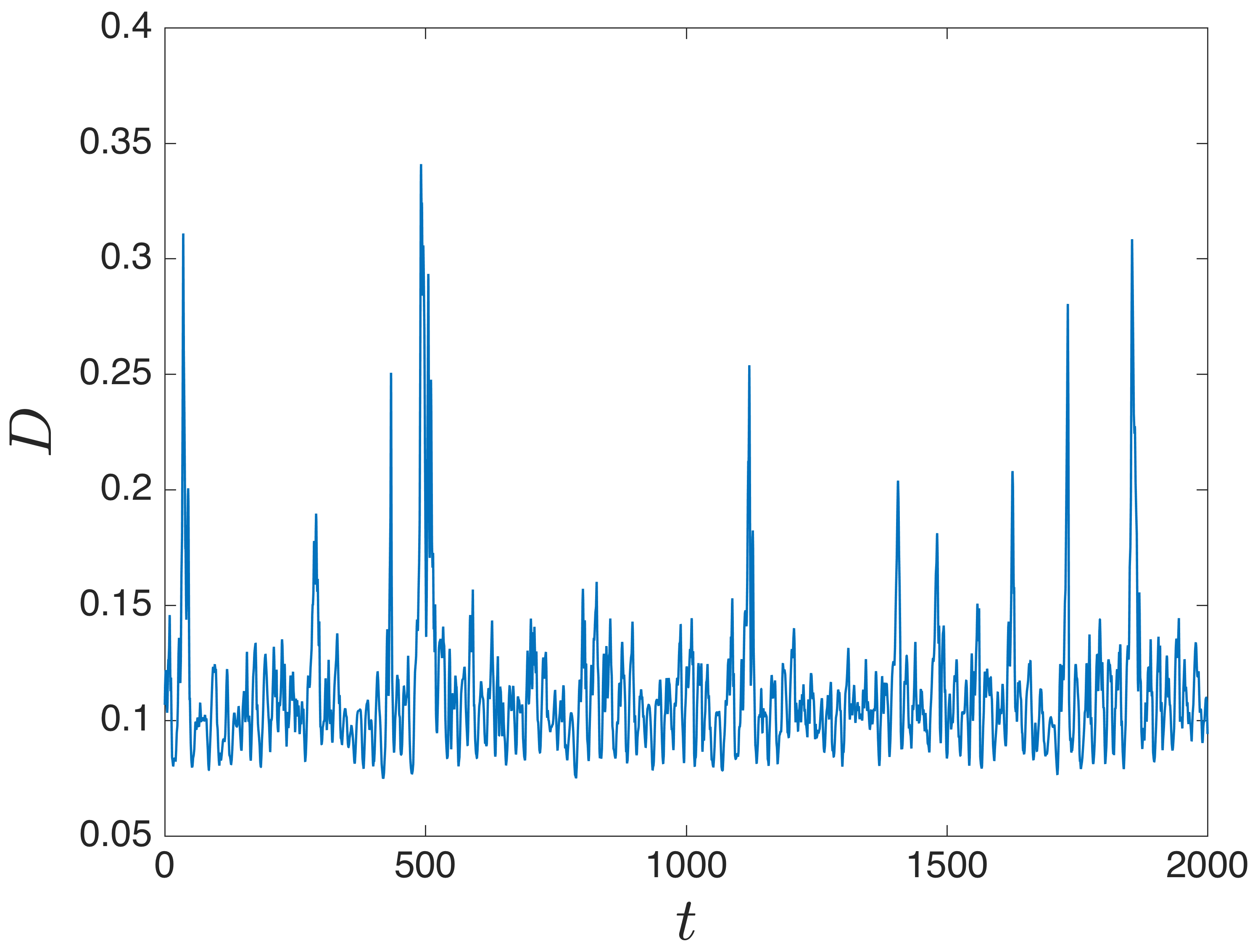}}\hspace{.1\textwidth}
\subfloat[]{\includegraphics[width=.4\textwidth]{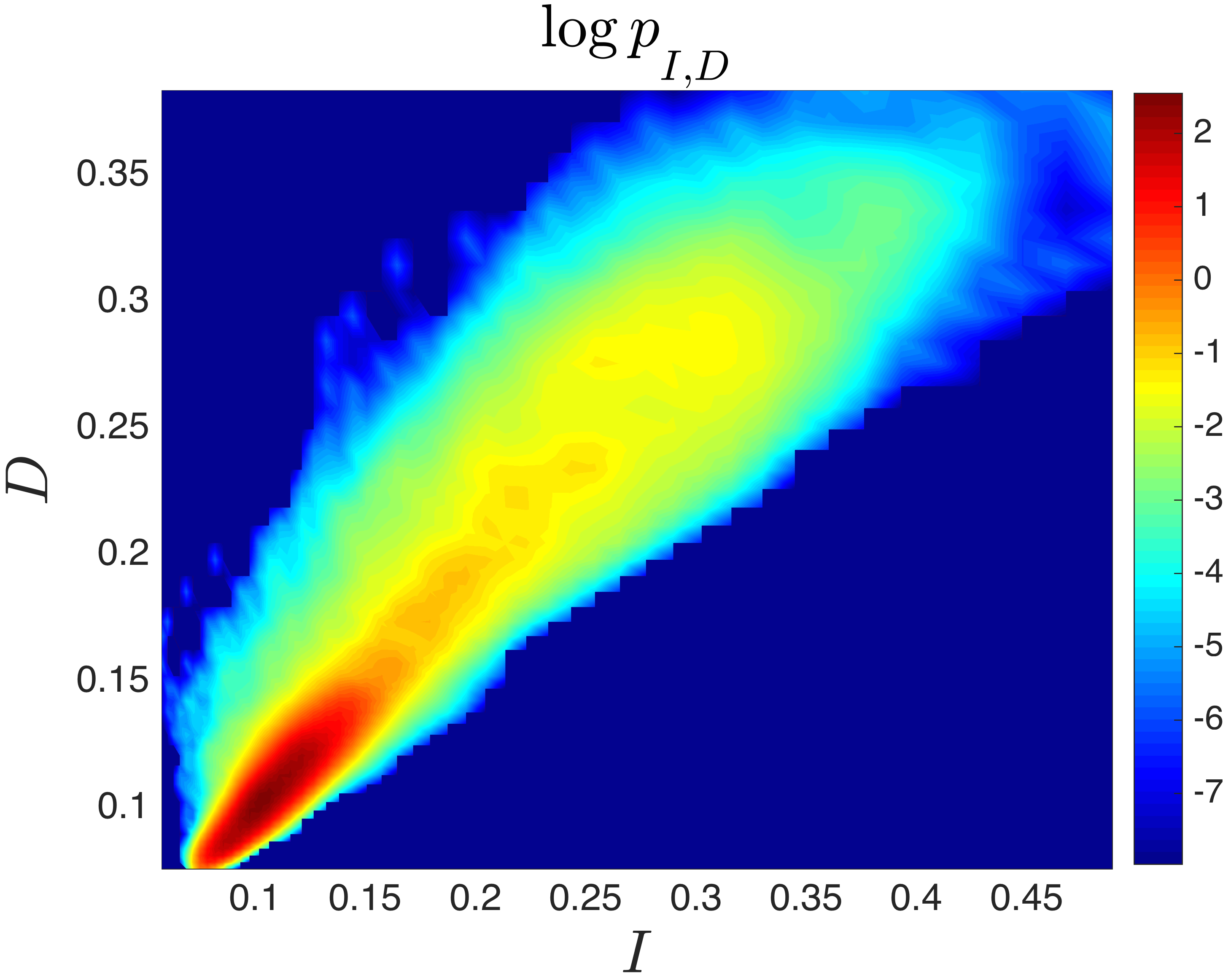}}
\caption{Intermittent bursts in the Kolmogorov flow at Reynolds number $Re=40$. (a) Time series of the energy dissipation rate $D$. 
(b) Logarithm of the joint probability density $\subs{p}{I,D}$ of the energy input rate $I$ and
the energy dissipation rate $D$.}
\label{fig:kolm_ts}
\end{figure*}

Because of the simplicity of the forcing $F$ and the boundary conditions, the \emph{Kolmogorov flow}
(i.e. the Navier--Stokes equations driven by the Kolmogorov forcing) has been studied extensively
both by numerical and analytical methods~\cite{obukhov1983,marchioro86,PlSiFi91, foias2001,CK13,faraz_adjoint}.
Similar variants of the Kolmogorov flow have also been investigated 
experimentally~\cite{batchaev1983,burgess1999,ouellette2008,Suri2017}.

In spite of the simplicity of the forcing and the boundary conditions, 
the Kolmogorov flow exhibits complex chaotic dynamics when the
Reynolds number $Re=\nu^{-1}$ is sufficiently large. In particular, 
the Kolmogorov flow is known to undergo intermittent bursts in this
chaotic regime~\cite{faraz_adjoint}. The bursts are detected by
monitoring certain system observables such as the 
energy dissipation rate $D:\funcspace\to \mathbb R^+$ and the energy input rate $I:\funcspace\to\mathbb R$,
\begin{equation}
D(u) = \frac{\nu}{(2\pi)^2}\int_{\mathbb T^2} |\nabla u |^2\id x,\quad 
I(u) = \frac{1}{(2\pi)^2}\int_{\mathbb T^2} u\cdot F\id x.
\label{eq:ID}
\end{equation}
The energy input rate $I$ measures the rate at which the external forcing 
pumps energy into the system. The energy dissipation rate $D$ measures 
the rate at which the system dissipates energy through diffusion.

Figure~\ref{fig:kolm_ts}(a) shows the 
time series of the energy dissipation rate along a typical trajectory of the
Kolmogorov flow at $Re=40$. This time series clearly exhibits 
chaotic, short-lived bursts. The bursts of the energy dissipation
are almost concurrent with the bursts of the energy input rate $I$. 
This can be inferred from figure~\ref{fig:kolm_ts}(b) showing the joint probability 
density $\subs{p}{I,D}$ associated with the joint probability distribution
\begin{equation}
\subs{F}{I,D}(I_0,D_0)=\Pmeas\left( u\in\funcspace : I(u)\leq I_0,\ D(u)\leq D_0\right),
\end{equation}
where $\Pmeas$ is the invariant probability measure induced by the solution map $\solmap{t}$
of the Kolmogorov flow (cf. section~\ref{sec:prelim}). In practice, 
the density $\subs{p}{I,D}$ is approximated from data sampled from long-time simulations 
along several trajectories~\cite{majda2005}.

Since the large values of $I$ correlate strongly with the large values of $D$ (figure~\ref{fig:kolm_ts}(b)), it is 
reasonable to assume that the same mechanism instigates the bursts of both quantities.
From a physical point of view, one is interested in the burst of the energy
dissipation rate $D$. However, since the energy input rate $I$ is linear in the velocity field
$u$, it is mathematically more convenient to work with this quantity. 

Given the simple form of the Kolmogorov forcing $F=\sin(k_fy)e_1$, 
the energy input rate~\eqref{eq:ID} can be written more explicitly as
$I(u(t))=-\mbox{Im}[a(0,k_f,t)]$ where $a(k_1,k_2,t)\in \mathbb C$ are the Fourier 
coefficients such that
\begin{equation}
u(x,t)=\sum_{k\in\mathbb Z^2}\frac{a(k,t)}{|k|}
\begin{pmatrix}
k_2\\
-k_1
\end{pmatrix}
e^{ik\cdot x},
\label{eq:fourier}
\end{equation}
where $k=(k_1,k_2)$. This Fourier series is written in a divergence-free form 
so that the incompressibility condition $\nabla \cdot u=0$ is ensured.
The energy input rate can be written in terms of the modulus $r(k,t)$ and phase
$\phi(k,t)$ of the Fourier coefficients as $I(u(t))=-r(0,k_f,t)\sin(\phi(0,k_f,t))$
where $a(k,t)=r(k,t)\exp(i\phi(k,t))$. Therefore, there are two scenarios through which 
the energy input rate $I$ can increase:
(i) For a fixed $r(0,k_f,t)$, the phase $\phi(0,k_f,t)$ approaches 
$-\pi/2$ resulting in $-\sin(\phi(0,k_f,t))\nearrow 1$ and subsequently
increasing $I$.
(ii) For a fixed phase $\phi(0,k_f,t)$, the modulus $r(0,k_f,t)$ increases
resulting in the growth of $I$.

Scenario (i) implies the alignment of the external forcing $F$ and the
velocity field $u(t)$ in the $L^2$ function space. This scenario, although appearing a priori more likely, is rejected based on
numerical observations (see Ref.~\cite{Farazmande1701533} for more details).
Instead, it is the increase in the modulus $r(0,k_f,t)$ that in turn leads to the
increase in $I$ during its bursts (scenario (ii)). The growth of $r(0,k_f,t)$
is only possible through the internal energy transfers operated by the nonlinear term
$u\cdot\nabla u$. 
It is known that the nonlinear term redistributes the energy (injected by the external forcing)
among the Fourier modes $a(k,t)$ in such a way that the total transfer of energy among modes is
zero~\cite{kraichnan1971,moffatt2014}.
Note that both the nonlinear term and the pressure gradient conserve energy
since 
\begin{equation}
\int_{\mathbb T^2} u\cdot \left(u\cdot\nabla u \right)\id x=0,\quad
\int_{\mathbb T^2} u\cdot\nabla p\, \id x=0.
\end{equation}
\begin{figure}
\centering
\subfloat[]{\includegraphics[width=.22\textwidth]{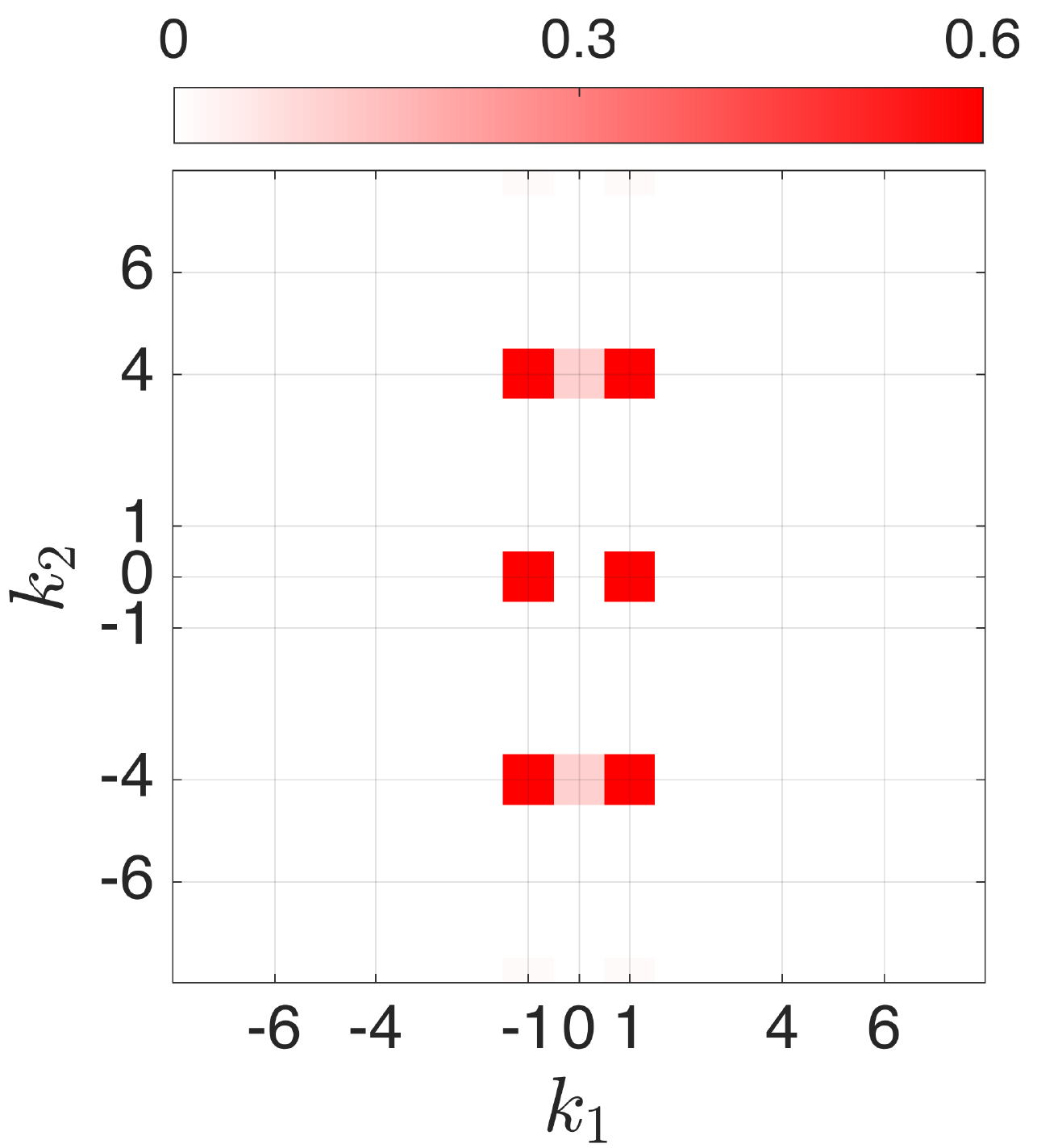}}\hspace{.04\textwidth}
\subfloat[]{\includegraphics[width=.22\textwidth]{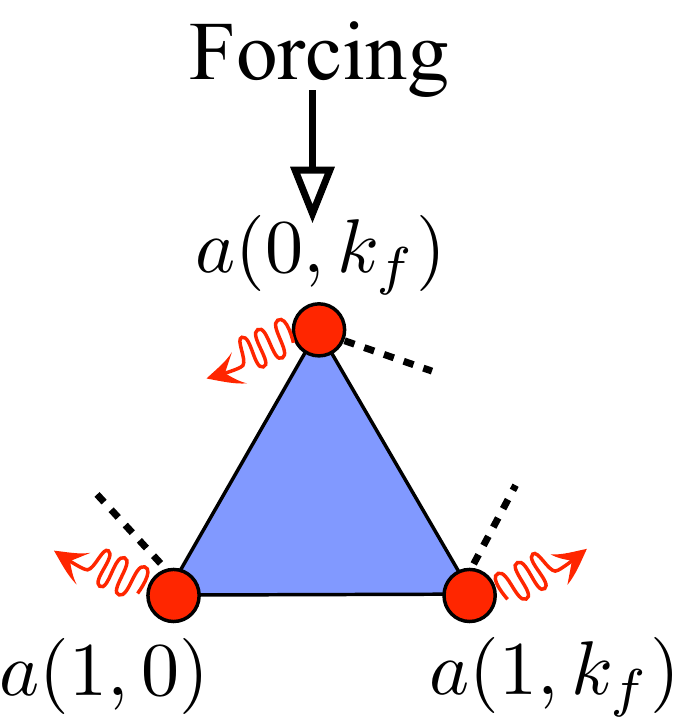}}
\caption{The optimal solution for the Kolmogorov flow at Reynolds number
$Re=40$ and forcing wave number $k_f=4$. (a) The optimal solution in 
the Fourier space. The color refers to the modulus of the Fourier modes, $|a(k_1,k_2)|$.
Most modes are vanishingly small (white color).
(b) A sketch of the main triad that is obtained from the optimal solution.
The other modes $(-1,0)$, $(1,k_f)$, etc. 
that are present in the optimal solution are repetitions of these three modes due to the 
complex conjugate relation $a(-k,t)=-a(k,t)^\ast$. The red wavy arrows represent the energy
dissipated by each mode. The dashed lines represent the coupling to other triads that not shown here.
}
\label{fig:kolm_optSol}
\end{figure}

Examining the structure of the Navier--Stokes in the Fourier space reveals that 
Fourier modes are coupled together in triads such that the mode $a(k,t)$ is
affected by pairs of modes $a(k',t)$ and $a(k'',t)$ with $k=k'+k''$~\cite{kraichnan1971}.
Each set of modes whose wavenumbers satisfy $k=k'+k''$ are referred to as a triad.
Since each mode may belong to several triads~\cite{moffatt2014}, they form a complex network of
triad interactions that continuously redistributes the energy among various modes.
As a result, it is not straightforward to discern the mode(s) responsible for the growth of 
the modulus of the mode $a(0,k_f)$, resulting in the bursts of the energy input $I$.

In Ref.~\cite{Farazmande1701533}, we employed a constrained optimization similar to~\eqref{eq:optGen2}
to discover the modal interactions that cause the extreme events in the Kolmogorov flow. Skipping the details, 
figure~\ref{fig:kolm_optSol} shows the obtained optimal solution in the Fourier space. 
This optimal solution essentially consists of three Fourier modes
with wavenumbers $(0,k_f)$, $(1,0)$ and $(1,k_f)$. Interestingly, these three modes form 
a triad since $(1,k_f)=(1,0)+(0,k_f)$. Moreover, the wavenumber $(0,k_f)$ is present in this triad
supporting scenario (ii) that postulated that the internal transfers of energy to mode $a(0,k_f)$
are responsible for extreme events in the Kolmogorov flow. 

\begin{figure}
\centering
\includegraphics[width=.45\textwidth]{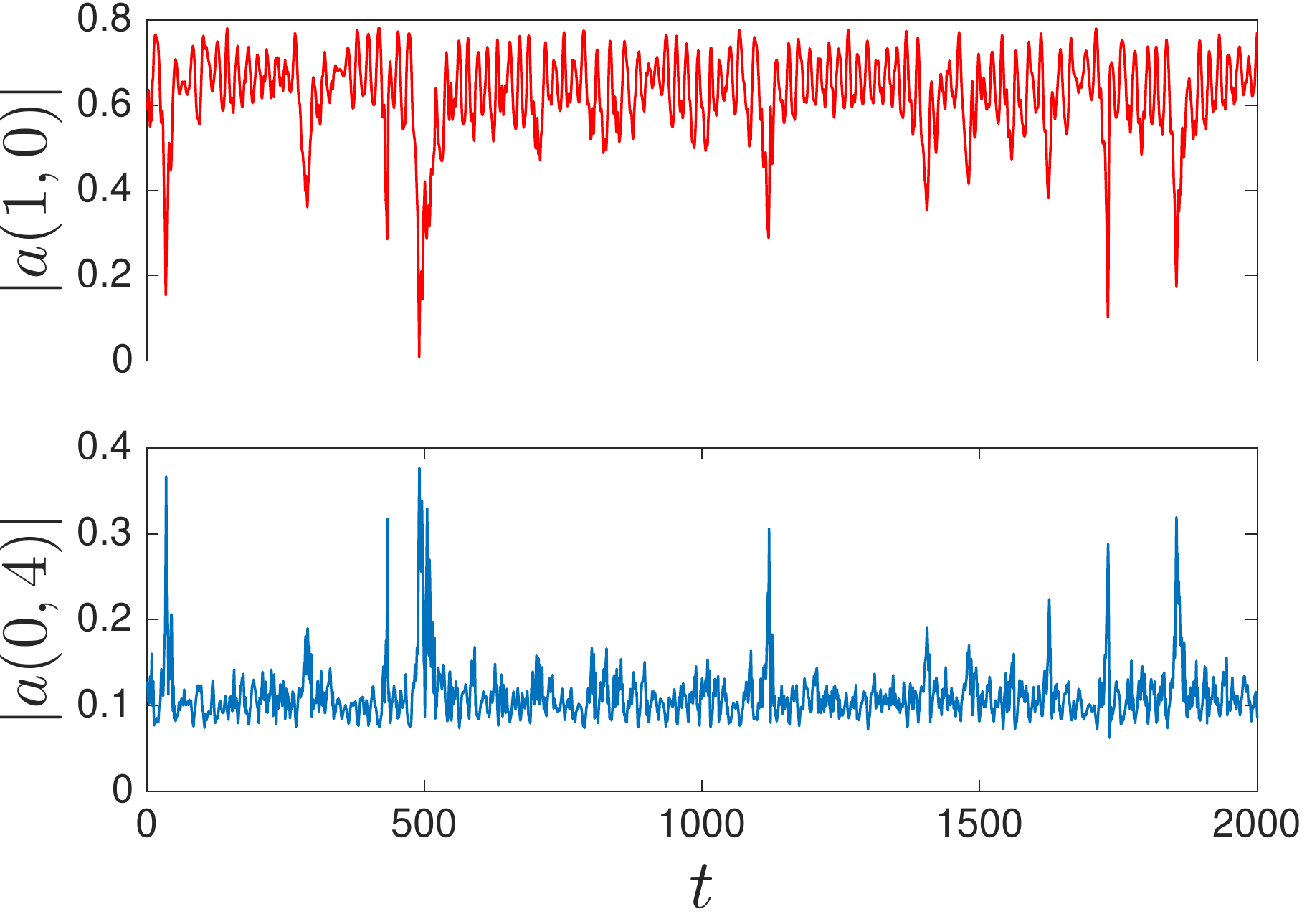}
\caption{Times series of the modulus of the modes $a(0,k_f)$
and $a(1,0)$ for the Kolmogorov flow at Reynolds number $Re=40$
and forcing wavenumber $k_f=4$. Note that $|a(0,k_f)|=r(0,k_f)$.}
\label{fig:kolm_pred_ts}
\end{figure}

Figure~\ref{fig:kolm_pred_ts} shows the evolution of the moduli
$|a(0,k_f)|$ and $|a(1,0)|$ along a typical trajectory of the Kolmogorov 
flow. First, we notice that $|a(0,k_f)|$ has bursts similar to those of 
the energy dissipation rate (see figure~\ref{fig:kolm_ts}). Secondly, 
the modulus $|a(1,0)|$ has sharp dips which are almost concurrent 
with the bursts of $|a(0,k_f)|$. This observation shows that, during
extreme events, the mode $a(1,0)$ loses its energy and transfers most of 
it to mode $a(0,k_f)$ through the triad interaction $(1,k_f)=(1,0)+(0,k_f)$.
The increase in $|a(0,k_f)|$, in turn, leads to an increase in the energy input
rate $I=-\mbox{Im}[a(0,k_f)]$ causing the observed bursts in $I$ (see figure~\ref{fig:kolm_ts}). 

How does this transfer of energy from a low wavenumber $(1,0)$ to a higher wavenumber
$(0,k_f)$ cause the bursts in the energy dissipation rate $D$? To answer this question we 
observe that 
\begin{equation}
D(u) = \nu\sum_{k\in\mathbb Z^2} |k|^2|a(k)|^2
\end{equation}
which follows directly from the definition of the energy dissipation~\eqref{eq:ID}
and the Fourier series~\eqref{eq:fourier}. The transfer of energy from the mode $a(1,0)$
to the mode $a(0,k_f)$ will significantly increase the energy dissipation rate 
since the term $|a(0,k_f)|^2$ is multiplied by a larger prefactor $k_f^2=16$
compared to the term $|a(1,0)|$ whose prefactor is $1$.

\subsection{Application to oceanic rogue waves}\label{sec:wave}
In this section, we consider the real-time prediction of rogue water waves. 
Rogue waves refer to unusually large waves when compared to the surrounding
waves. While there is no rigorous definition of a rogue wave, it is customary to 
define it as a wave whose height exceeds twice the significant wave height.
For a given sea state, the significant wave height refers 
to four times the standard deviation of the surface elevation~\cite{dysthe08}.

\begin{figure}
\centering
\includegraphics[width=.45\textwidth]{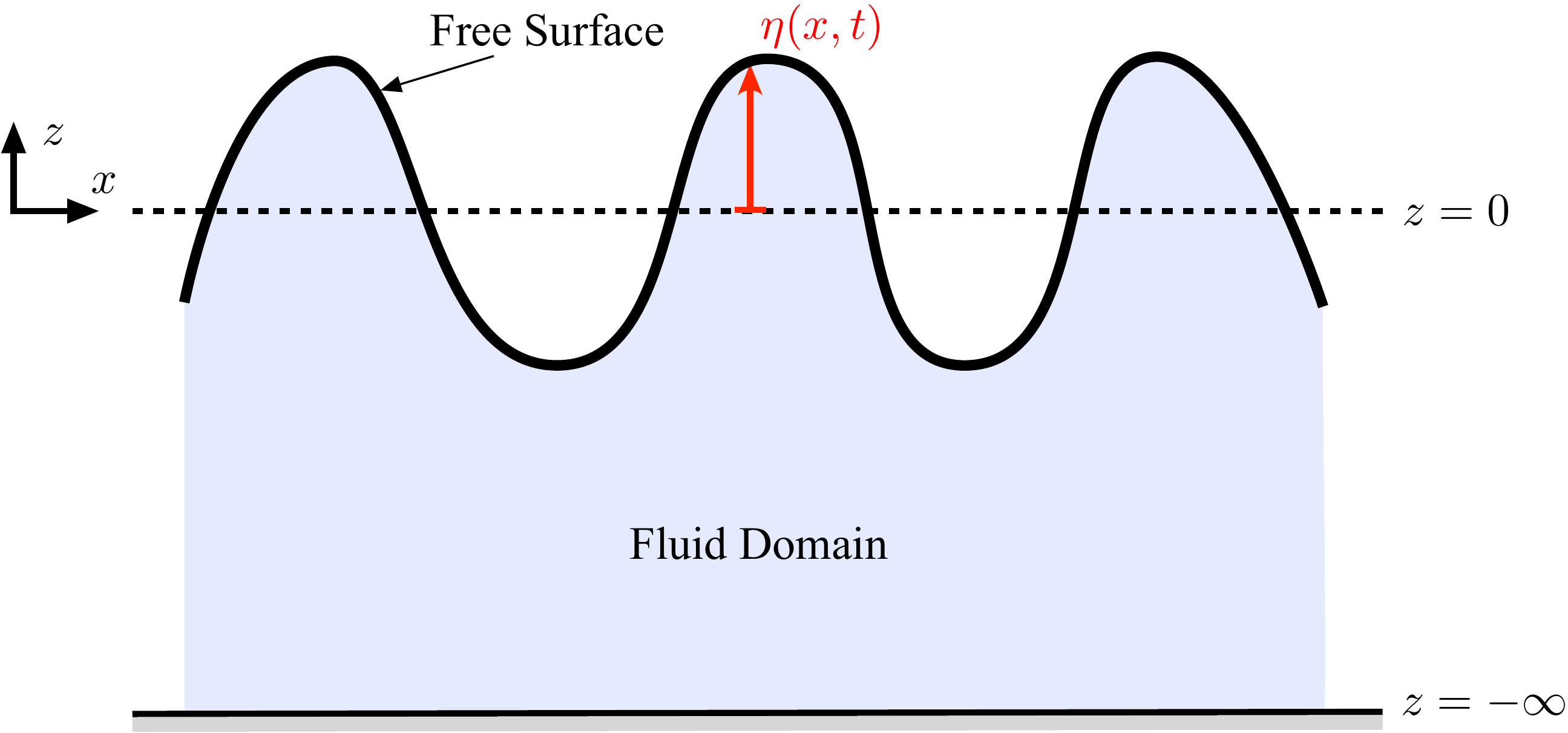}
\caption{A sketch of the water wave problem. At any time $t$, the free surface $z=\eta(x,t)$ is given as a graph over the horizontal coordinate $x$.}
\label{fig:wave_1d}
\end{figure}

As a starting point, we consider the free-surface, as an unidirectional, irrotational flow in deep seas. 
The surface elevation $\eta:(x,t)\mapsto \eta(x,t)$ is a function of the horizontal spatial variable $x$ and time $t$ (see figure~\ref{fig:wave_1d}).
The vertical coordinates are denoted by the variable $z$, such that the 
velocity potential is given by $\phi: (x,z,t)\mapsto \phi(x,z,t)$.
In this setting, the water waves are governed by the set of equations~\cite{stoker},
\begin{subequations}
\label{eq:ww_eq}
\begin{equation}
\frac{\partial \phi}{\partial t}+\frac12 |\nabla \phi|^2+g z=0,\quad z=\eta (x,t),
\label{eq:bernoulli}
\end{equation}
\begin{equation}
\Delta \phi = 0,\quad -\infty <z<\eta (x,t),
\label{eq:div-free}
\end{equation}
\begin{equation}
\frac{\partial \phi}{\partial z}  = 0,\quad  z=-\infty,
\label{eq:bc_bottom}
\end{equation}
\begin{equation}
\frac{\partial \eta }{\partial t}+\frac{\partial \phi }{\partial x}\frac{\partial \eta }{\partial x}-\frac{\partial \phi }{\partial z}=0,\quad
z=\eta(x,t).
\label{eq:bc_top}
\end{equation}
\end{subequations}
Equation~\eqref{eq:bernoulli} is the Bernoulli equation for irrotational flows with a free surface.
Equation~\eqref{eq:div-free} follows from the conservation of mass.
Equations~\eqref{eq:bc_bottom} and~\eqref{eq:bc_top} are the
boundary conditions at the bottom of the sea and the surface, respectively.
The constant $g$ denotes the gravitational acceleration.

To solve the water wave equations~\eqref{eq:ww_eq} numerically, we need the initial surface elevation $\eta(x,0)$
and the initial velocity potential $\phi(x,z,0)$. While the practical measurement of the surface elevation is possible~\cite{nieto04, fu_2011,story11,borge13}, 
measuring the entire velocity potential beneath the surface remains a challenging task. 
Therefore, it is highly desirable to decouple the surface evolution from the velocity potential. 
This motivates the use of the so-called \emph{envelope equations}, an approximation to the water wave equations
that only involves the surface elevation $\eta$. 
\begin{figure}
\centering
\subfloat[]{\includegraphics[width=.24\textwidth]{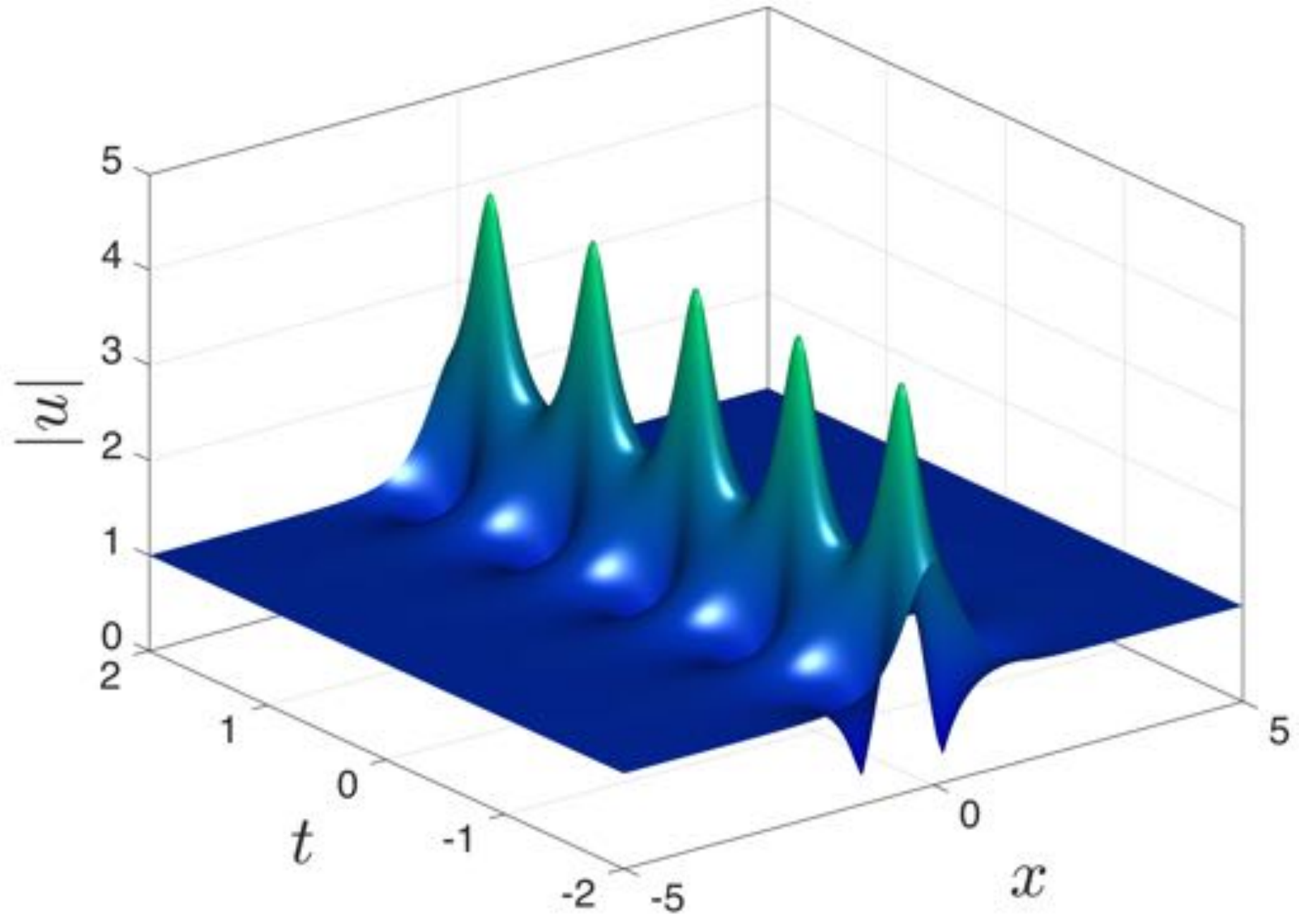}}
\subfloat[]{\includegraphics[width=.24\textwidth]{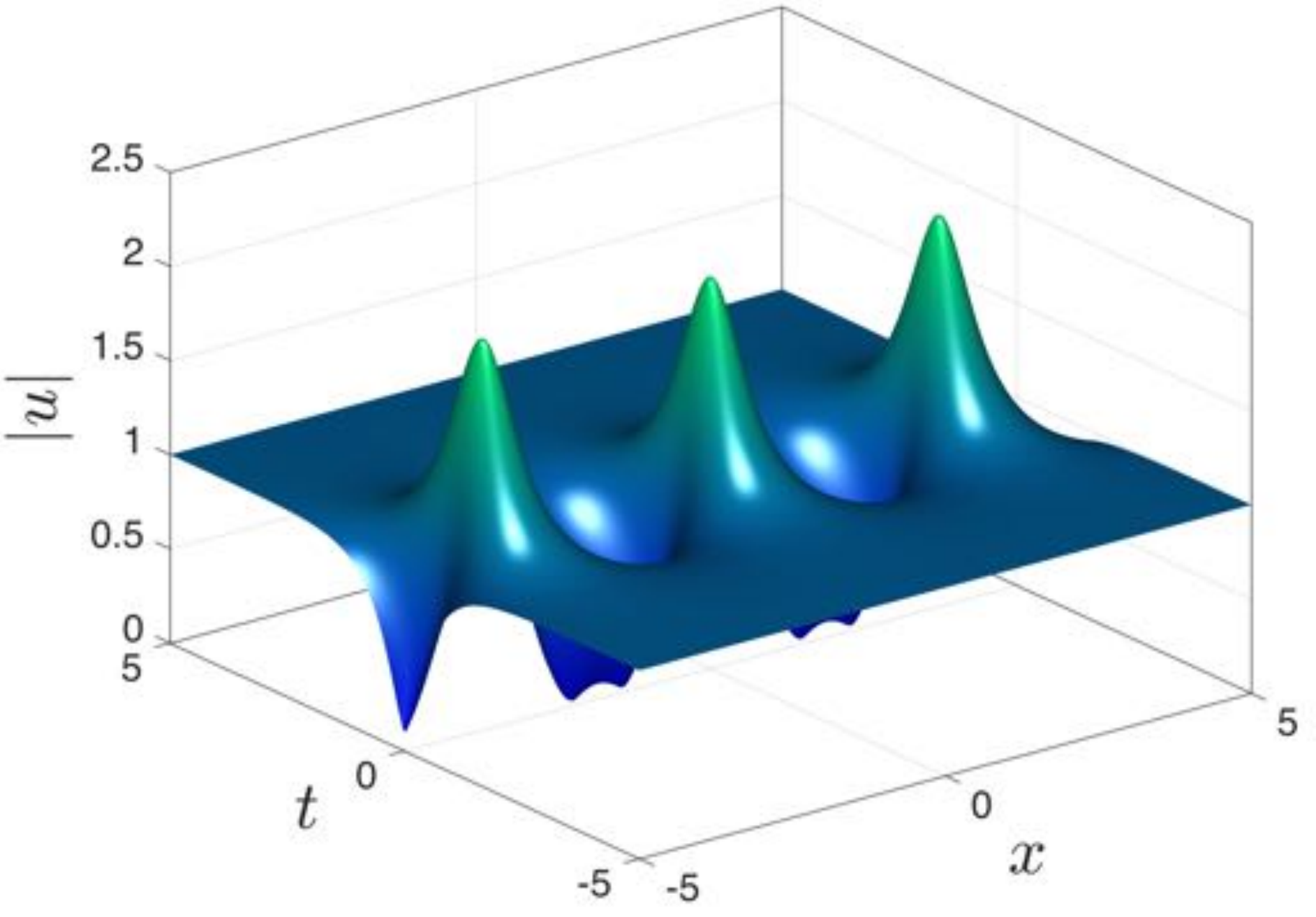}}\\
\subfloat[]{\includegraphics[width=.24\textwidth]{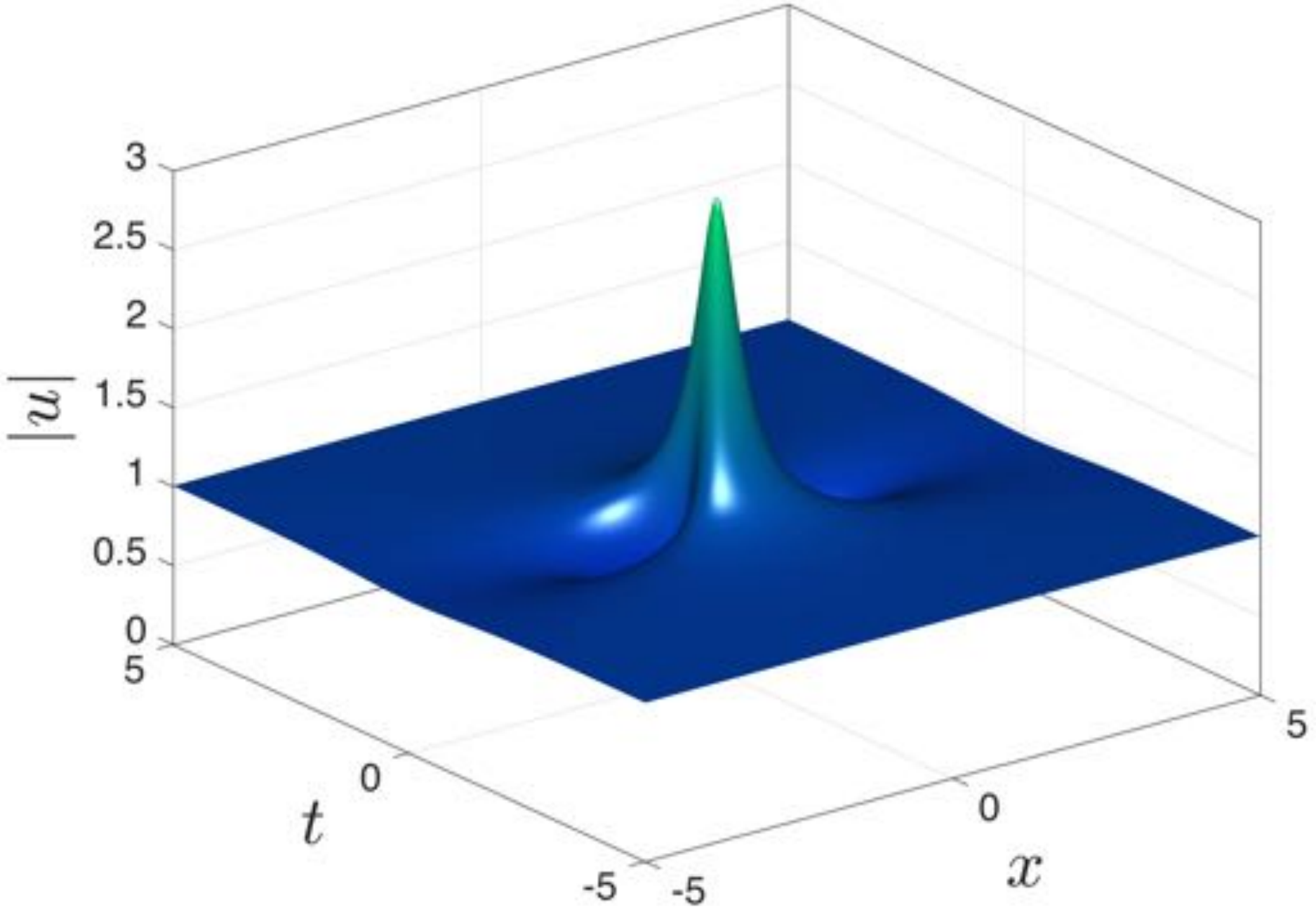}}
\caption{Breather solutions of the NLS equation.
(a) The Ma breather~\cite{Ma1979} is periodic in time and localized in space.
(b) The Akhmediev breather~\cite{Akhmediev1986} is localized in time and periodic in space.
(c) The Peregrine breather~\cite{peregrine83} is doubly localized in both space and time.}
\label{fig:breathers}
\end{figure}

The envelope equations govern the evolution of perturbations to the linear waves
\begin{equation}
\phi = \frac{ag}{\omega_0}e^{k_0z}\sin(k_0x-\omega_0t),\quad \eta = a\cos(k_0x-\omega_0t),
\label{eq:linwave}
\end{equation}
where $a\geq 0$ is the wave amplitude, $\lambda_0=2\pi k_0^{-1}$ is the wavelength and 
$T_0=2\pi \omega_0^{-1}$ is the wave period. The pair~\eqref{eq:linwave} is referred to as
a linear wave since it is an exact solution of the linear part of the water wave equation~\eqref{eq:ww_eq}. 
The wave frequency $\omega_0$ and the
wave number $k_0$ satisfy the linear dispersion relation $\omega_0(k_0)=\sqrt{gk_0}$.
The envelope equations describe the evolution of small perturbations
to the linear wave~\eqref{eq:linwave}. These perturbations are of the modulation form
\begin{equation}
\eta(x,t)=\mbox{Re}\left\{u(x,t)e^{i(k_0x-\omega_0t)}\right\},
\end{equation}
where $u\in \mathbb C$ is 
the complex wave envelope. Perturbation analysis shows that, to the first order, $u(x,t)$ satisfies the
nonlinear Schr\"odinger (NLS) equation~\cite{benney1967,zakharov68,Hasimoto1972},
\begin{equation}
\frac{\partial u}{\partial t}+\frac12 \frac{\partial u}{\partial x} + \frac{i}{8}\frac{\partial^2 u}{\partial x^2}+\frac{i}{2}|u|^2 u =0,
\end{equation}
where we have normalized the space and time variables with the wavelength and 
wave period of the underlying periodic wave train so that $x\mapsto k_0x$ and
$t\mapsto \omega_0 t$. 
This perturbation analysis is valid under certain assumptions~\cite{Hasimoto1972,yuen1975}, including that
the wave steepness $\epsilon=ak_0$ is small, i.e., $\epsilon\ll 1$.
These assumptions can be relaxed by considering higher-order terms in the 
perturbation analysis~\cite{zakharov68,dysthe79,trulsen2000}.

Several exact solutions of the NLS equation have been found over the years. Figure~\ref{fig:breathers}
shows three types of the so-called \emph{breather} solutions of the NLS equation. These solutions
are localized in time or space or both. Of particular interest to us 
is the Peregrine breather (panel c) since it mimics the rogue waves in the sense that it starts from a plane wave, develops into a
localized large wave and again decays to a plane wave. 

The breather solutions have been observed in carefully controlled experiments~\cite{yuen1975,chabchoub2011,chabchoub2012a,chabchoub2012b,narhi2016}.
However, real ocean waves are irregular wave fields consisting of many dispersive wave groups so that
the detection of breathers from a given wave field becomes a difficult task~\cite{chabchoub2016}.
More importantly, these exact breather solutions are not the only possible mechanism for rogue wave formation. 
For instance, Cousins and Sapsis~\cite{cousins15} studied the evolution of initial wave groups of the
form $|u|=A_0\sinh(x/L_0)$ for various combinations of wave amplitude $A_0$ and length scale $L_0$.
They find a range of parameters $(A_0,L_0)$ where the initially small wave groups develop into a
rogue wave at a later time when evolved under the NLS equation. 
\begin{figure}
\centering
\includegraphics[width=.48\textwidth]{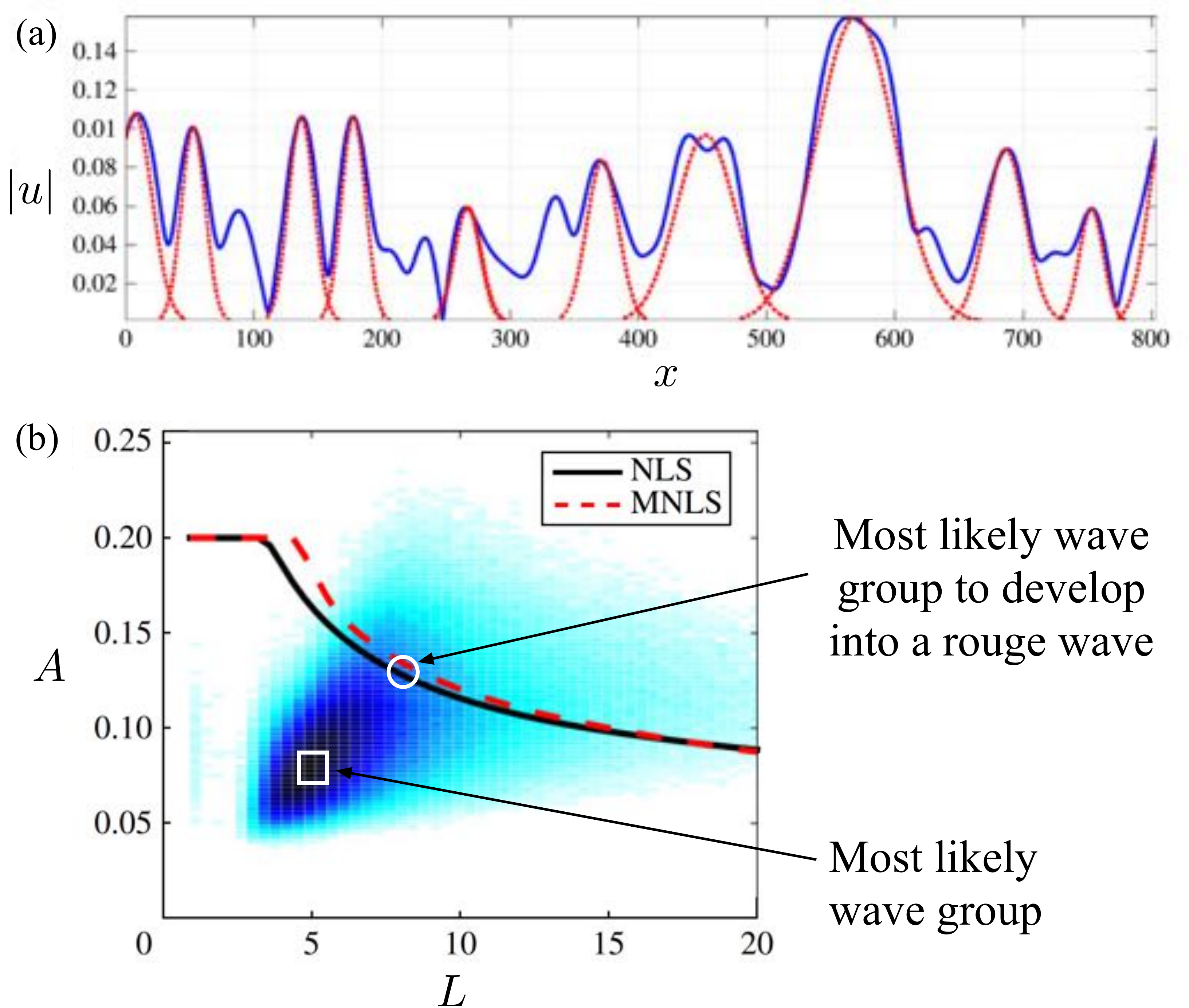}
\caption{Reduced-order prediction of rogue waves.
(a) An irregular wave field (solid blue) is approximated by the superposition of localized wave groups (dotted red).
(b) The joint probability density function of the length scales $L$ and amplitudes $A$ obtained from 
decomposing many realizations of random wave fields into localized wave groups. Darker colors mark higher probability. 
The solid black curve mark the boundary above which the wave groups develop into a rogue wave at some point in the future
when evolved under the NLS equation. The dashed red curve marks the same boundary but under the modified NLS (MNLS) equation~\cite{dysthe79}.
Figure reproduced from Ref.~\cite{cousins16}.}
\label{fig:cs2016}
\end{figure}

One can approximate an irregular wave field as a superposition of 
localized wave groups with $\sech$ envelopes,
\begin{equation}
|u(x)| \simeq \sum_{i=1}^n A_i\sech \left(\frac{x-x_i}{L_i}\right),
\label{eq:sech}
\end{equation}
where the parameters $(A_i,L_i,x_i)$ are chosen so that the approximation error is minimized.
An example of such a decomposition is shown in figure~\ref{fig:cs2016}(a). 
Figure~\ref{fig:cs2016}(b) shows the joint probability density function (PDF) of the parameters $(A_i,L_i)$.
This PDF is computed by approximating many realizations of random waves
with the superposition~\eqref{eq:sech}.

This joint PDF contains several interesting pieces of information. In particular, 
it indicates the most likely combination of length scale and amplitude of wave groups
in a given random sea (marked with a white square). These wave groups, however, do not necessarily develop into 
rogue waves. The solid black curve in figure~\ref{fig:cs2016}(b) marks the boundary between wave groups that develop into 
a rogue wave at some point in the future (the wave groups above the curve) and those that do not (the wave groups below the curve).
The intersection of this curve with the joint PDF determines the most `dangerous' waves (marked by a white circle), i.e., 
the most likely wave groups that will develop into a 
rogue waves at some point in the future. 

Cousins and Sapsis~\cite{cousins16} used this information to develop a reduced-order method for 
prediction of rogue waves in unidirectional water waves in deep sea. This method does not require the numerical integration of the NLS equations and, as a result, is
computationally much less expensive. In addition, the reduced-order model only requires the knowledge of 
localized wave groups that form the wave field. As such, this method can be applied to cases where the wave field is only partially known or when the measurement 
resolution is low. Later, Farazmand and Sapsis~\cite{Farazmand2017} generalized the reduced-order prediction of rogue waves to 
two-dimensional water waves.

We finally point out that the most `dangerous' wave groups could alternatively be found as solutions to a constrained optimization
problem similar to~\eqref{eq:opt_general}. However, since the computational cost of generating figure~\ref{fig:cs2016} is not 
prohibitive, the most dangerous waves were estimated directly from the joint PDF. Later, in the context of large deviation theory, Dematteis et al.
\cite{Dematteis2018} obtained similar results by solving a constrained optimization problem.

\section{Prediction of extreme events}\label{sec:prediction}
\begin{figure}
\centering
\includegraphics[width=.45\textwidth]{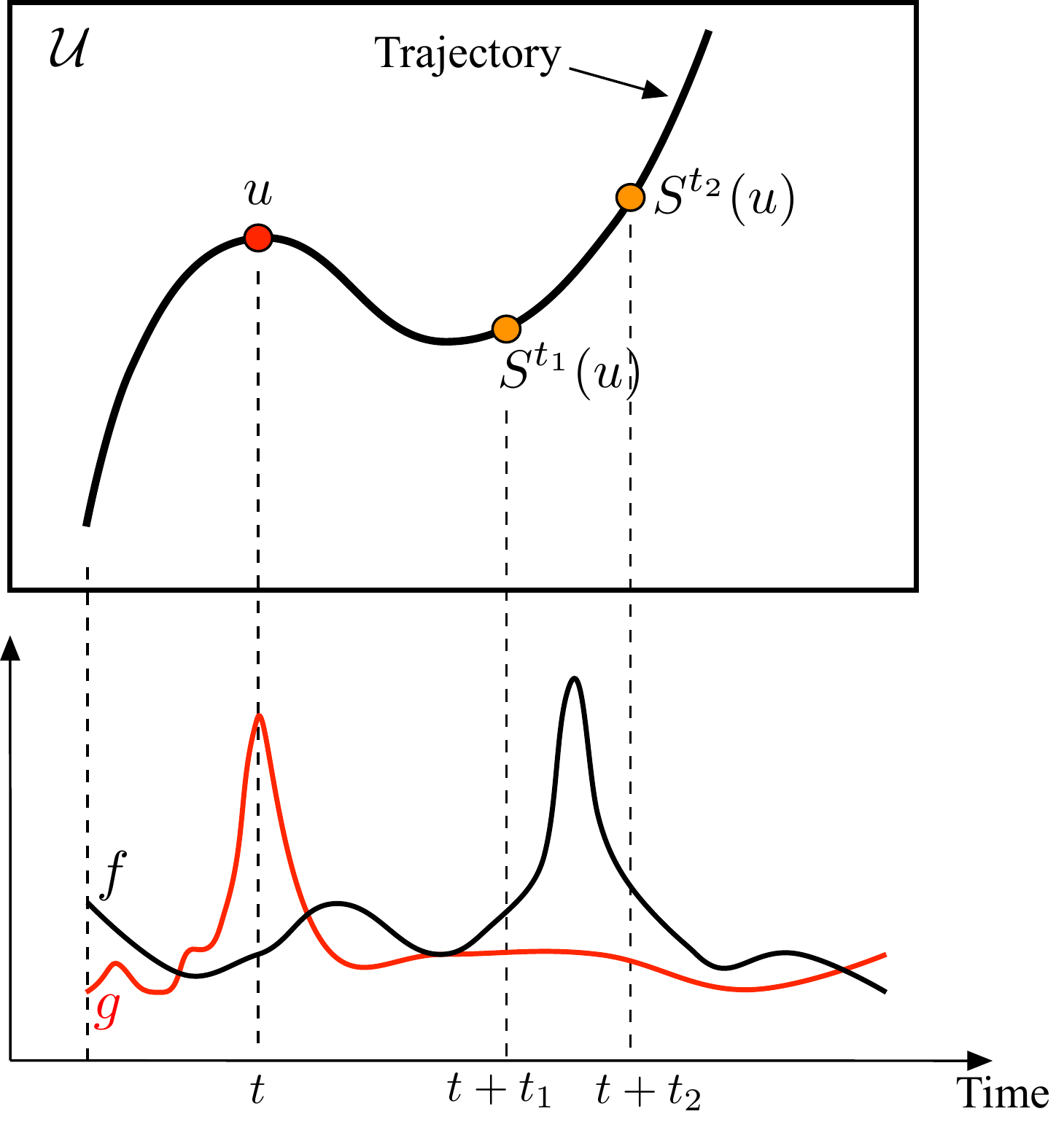}
\caption{The evolution of the observable $f$ and the indicator $g$ along a
trajectory in the state space $\funcspace$. The large value of 
the observable $g$ at a state $u$ signals an upcoming
large value of the observable $\obs$ over the future time interval $[t_1,t_2]$.}
\label{fig:pred_traj}
\end{figure}

In this section, we turn our attention to the discovery of indicators of extreme events. 
Given an observable $\obs:\funcspace\to\mathbb R$ of the system~\eqref{eq:masterEq}, we seek indicators 
$g:\funcspace\to \mathbb R$ whose evolution along a trajectory $u(t)$ signal
an upcoming extreme value of the observable $\obs$. This is sketched in figure~\ref{fig:pred_traj}
where the indicator $g$ attains a relatively large value at time $t$ just before the observable $\obs$
attains a large value over the future time interval $[t+t_1,t+t_2]$. Note that the indicator $g$ is itself 
an observable of the system, but it is carefully chosen such that it predicts
the extreme events associated with $\obs$.

As a first step, we need to quantify the predictive skill of an observable $g$.
To this end, we define a number of quantities. In particular, we define the maximum 
observable values over a future time interval,
\begin{equation}
f_m(u;t_1,t_2)=\max_{t_1\leq \tau\leq t_2}\obs\left(\solmap{\tau}(u)\right),
\label{eq:fm}
\end{equation}
where $u\in\funcspace$ is a state and $0<t_1\leq t_2$. We refer to $t_1$ as the time 
horizon of the prediction. In the special case where $t_1=t_2$, we have 
\begin{equation}
\obs_m(u;t_1,t_1)=f(\solmap{t_1}(u)),
\end{equation}
where $f_m(u;t_1,t_1)$ is the value of the observable at $t_1$ time units in the future
if the current state of the system is $u$. 
If $t_1\neq t_2$, then $\obs_m(u;t_1,t_2)$ is the maximum of the 
observable $\obs$ over the future time interval $[t_1,t_2]$ along the trajectory passing through the state $u$.
Our goal therefore is to find an indicator $g:\funcspace\to \mathbb R$
whose large values correlate strongly with the large values of $f_m(\cdot;t_1,t_2)$ for appropriate 
choices of $t_1$ and $t_2$. We quantify this correlation through conditional statistics.

\subsection{Conditional statistics for extreme events}\label{sec:cond_stat}

\begin{figure*}
\centering
\includegraphics[width=.75\textwidth]{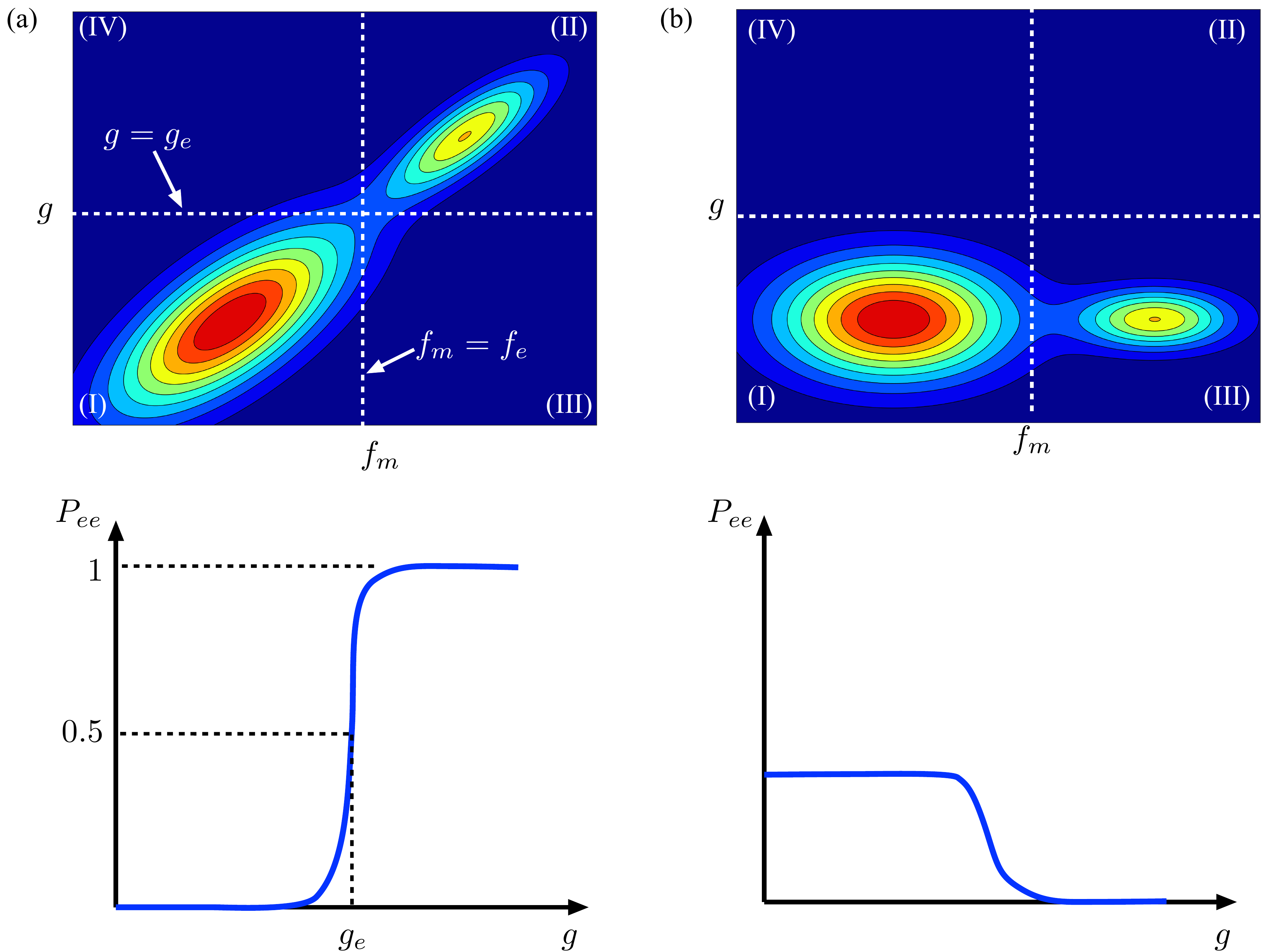}
\caption{Two possible conditional PDFs for the predictor $g(t)$ of a future extreme event $f_m(t;t_1,t_2)=\max_{\tau\in[t+t_1,t+t_2]}f(\tau)$ 
of an observable $f$. (a) A skillful predictor characterized by low false positives and low false negatives.
(b) A `bad' predictor that returns high false negatives. (c) A `bad' predictor that returns high false positives and high false negatives.}
\label{fig:condPDF}
\end{figure*}

Let $\subs{p}{f_m,g}\in L^1(\mathbb R^2)$
denote the probability density associated with the joint probability distribution,
\begin{align}
\subs{F}{f_m,g}(f_0,g_0) & = \Pmeas\left(u\in\funcspace : f_m(u;t_1,t_2)\leq f_0,g(u)\leq g_0\right)\nonumber\\
& =\int_{-\infty}^{f_0}\int_{-\infty}^{g_0}\subs{p}{f_m,g}(a,b)\id b\,\id a,
\end{align}
for given $0<t_1<t_2$. Similarly, let $\subs{p}{g}\in L^1(\mathbb R)$ denote the 
probability density associated with the distribution 
\begin{align}
F_g(g_0)&=\Pmeas(u\in\funcspace: g(u)\leq g_0)\nonumber\\
& = \int_{-\infty}^{g_0} \subs{p}{g}(b)\id b.
\end{align}
Therefore, the conditional probability density $\subs{p}{f_m|g}$ is given by
\begin{equation}
\subs{p}{f_m|g}=\frac{\subs{p}{f_m,g}}{\subs{p}{g}}.
\label{eq:condPDF}
\end{equation}
Roughly speaking, the conditional probability density $\subs{p}{f_m|g}(f_0,g_0)$ measures the probability of $f_m(u;t_1,t_2)=f_0$ given that 
$g(u)=g_0$.

Recall from Definition~\ref{def:ee} that an extreme event corresponds to $\obs>\obs_e$.
Therefore, an extreme event takes place over the future time interval $[t_1,t_2]$
if $f_m(u;t_1,t_2)>f_e$. An ideal indicator $g$ of extreme events should have a 
corresponding threshold $g_e$ such that $g(u)>g_e$ implies 
$f_m(u;t_1,t_2)>f_e$. Conversely, $g(u)<g_e$ indicates that no upcoming 
extreme events are expected, that is
$f_m(u;t_1,t_2)<f_e$. The corresponding conditional PDF $\subs{p}{f_m|g}$ of
such an ideal indicator is shown in figure~\ref{fig:condPDF}(a).
Unsuccessful predictions correspond to the cases where either 
$\{g(u)>g_e\ \mbox{and}\ f_m(u;t_1,t_2)<f_e\}$ or $\{g(u)<g_e\ \mbox{and}\ f_m(u;t_1,t_2)>f_e\}$.

These four possibilities are summarized below:
\begin{enumerate}[label=(\Roman*)]
\item Correct Rejections: $g<g_e$ and $\obs_m<\obs_e$.

The indicator correctly predicts that no extreme events are upcoming.
\item Correct Predictions: $g>g_e$ and $\obs_m>\obs_e$.

The indicator correctly predicts an upcoming extreme event. 
\item False Negatives: $g<g_e$ and $\obs_m>\obs_e$.

The indicator fails to predict an upcoming extreme event.
\item False Positives: $g>g_e$ and $\obs_m<\obs_e$.

The indicator falsely predicts an upcoming extreme event.
\end{enumerate}

These possibilities divide the conditional PDF plots of $\subs{p}{f_m|g}$ 
into four quadrants (see figure~\ref{fig:condPDF}). Figure~\ref{fig:condPDF}(a)
sketches the conditional PDF corresponding to a reliable indicator: there is a threshold $g_e$ for
which negligible false positives and false negatives are recorded (low density in quadrants III and IV). 
Figure~\ref{fig:condPDF}(b), on the other hand, sketches an unreliable predictor. For this indicator 
there is no choice of the threshold $g_e$ that leads to negligible amount of false positives and false negatives. 
The sketched threshold, for instance, returns no false positives but at the same time, 
does not predict any of the extreme events, hence, returning high false negatives.

The conditional PDF $\subs{p}{f_m|g}$ also enables us to quantify the probability
that an extreme event will take place over the future time interval $[t_1,t_2]$, 
given the value of the indicator at the present time. More precisely, we can measure the probability
that $\obs_m(u;t_1,t_2)>\obs_e$, given that $g(u) = g_0$. We refer to this quantity as
the \emph{probability of upcoming extreme events} (or probability of extreme events, for short).

\begin{defn}[Probability of Upcoming Extremes]
For a given observable $\obs:\funcspace\to\mathbb R$, its associated future maximum 
$f_m(\cdot;t_1,t_2):\funcspace\to\mathbb R$
and an indicator $g:\funcspace\to \mathbb R$, we define the 
probability of an upcoming extreme event as
\begin{equation}
P_{ee}(g_0)=\int_{f_e}^\infty \subs{p}{f_m|g}(a,g_0)\id a,
\end{equation}
where $\subs{p}{f_m|g}$ is the conditional PDF defined in~\eqref{eq:condPDF}
and $\obs_e$ is the threshold of extreme events (see Definition~\ref{def:ee}).
\label{def:Pee}
\end{defn}

Roughly speaking, in terms of the invariant probability measure $\Pmeas$, $P_{ee}(g_0)$
measures
\begin{equation}
\Pmeas\left( u\in\funcspace: \obs_m(u;t_1,t_2)>f_e\; |\; g(u) = g_0\right). 
\end{equation}
For a reliable indicator $g$, we have $P_{ee}(g_0)\simeq 0$ if $g_0<g_e$ 
and $P_{ee}(g_0)\simeq 1$ if $g_0>g_e$, with a sharp transition in between (see figure~\ref{fig:condPDF}).

\subsection{Applications}\label{sec:pred_app}
Now we demonstrate how these quantities are applied in practice by returning to the examples 
discussed in sections~\ref{sec:kolm} and~\ref{sec:wave}. Recall from section~\ref{sec:kolm} that 
the extreme events in the Kolmogorov flow (i.e., large values of the energy dissipation rate) occur when a significant amount of 
energy is transfered from the mode $a(1,0)$ to the forcing mode $a(0,k_f)$.
As a result, during the extreme events, the mode $a(1,0)$ loses energy, resulting in
relatively small values of $|a(1,0)|$. Visual examination of the time series of the energy dissipation rate $D$
and modulus $|a(1,0)|$ suggest that this energy loss takes place shortly before the extreme values of 
the energy dissipation rate are registered. This observation suggests that small values of $|a(1,0)|$ can
be used for short-term prediction of the extreme events. 

\begin{figure*}
\centering
\subfloat[]{\includegraphics[width=.3\textwidth]{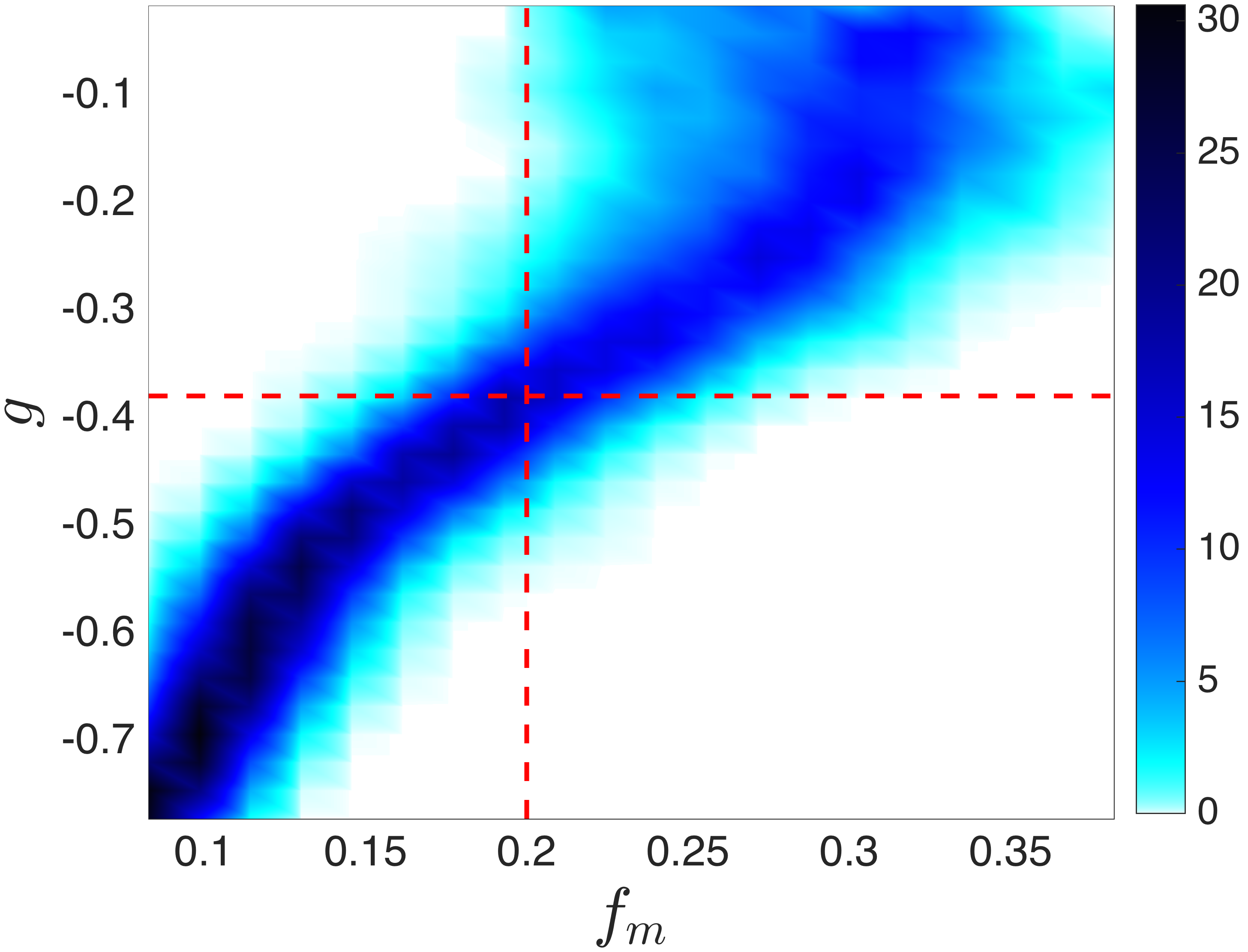}}\hspace{.04\textwidth}
\subfloat[]{\includegraphics[width=.3\textwidth]{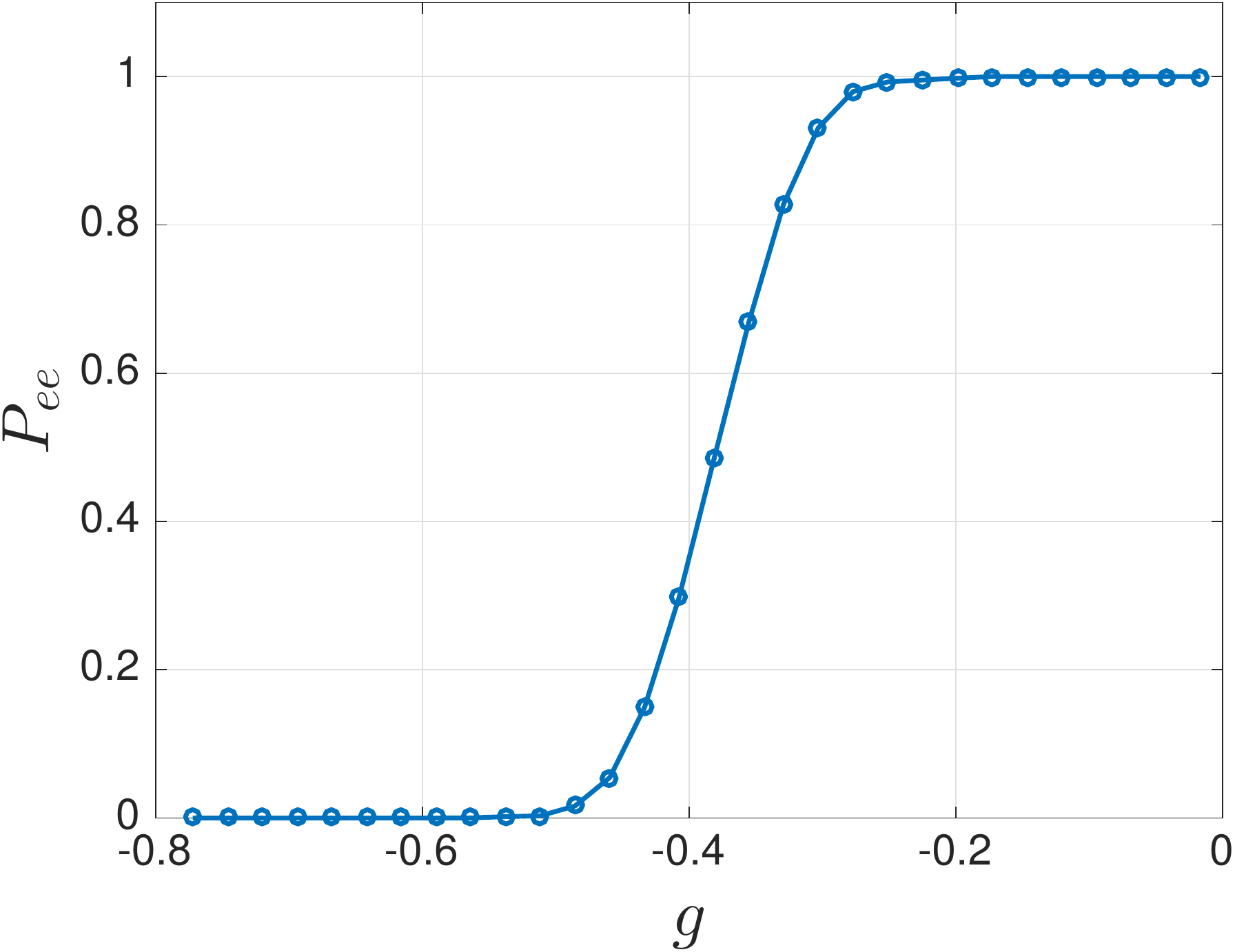}}\hspace{.04\textwidth}
\subfloat[]{\includegraphics[width=.3\textwidth]{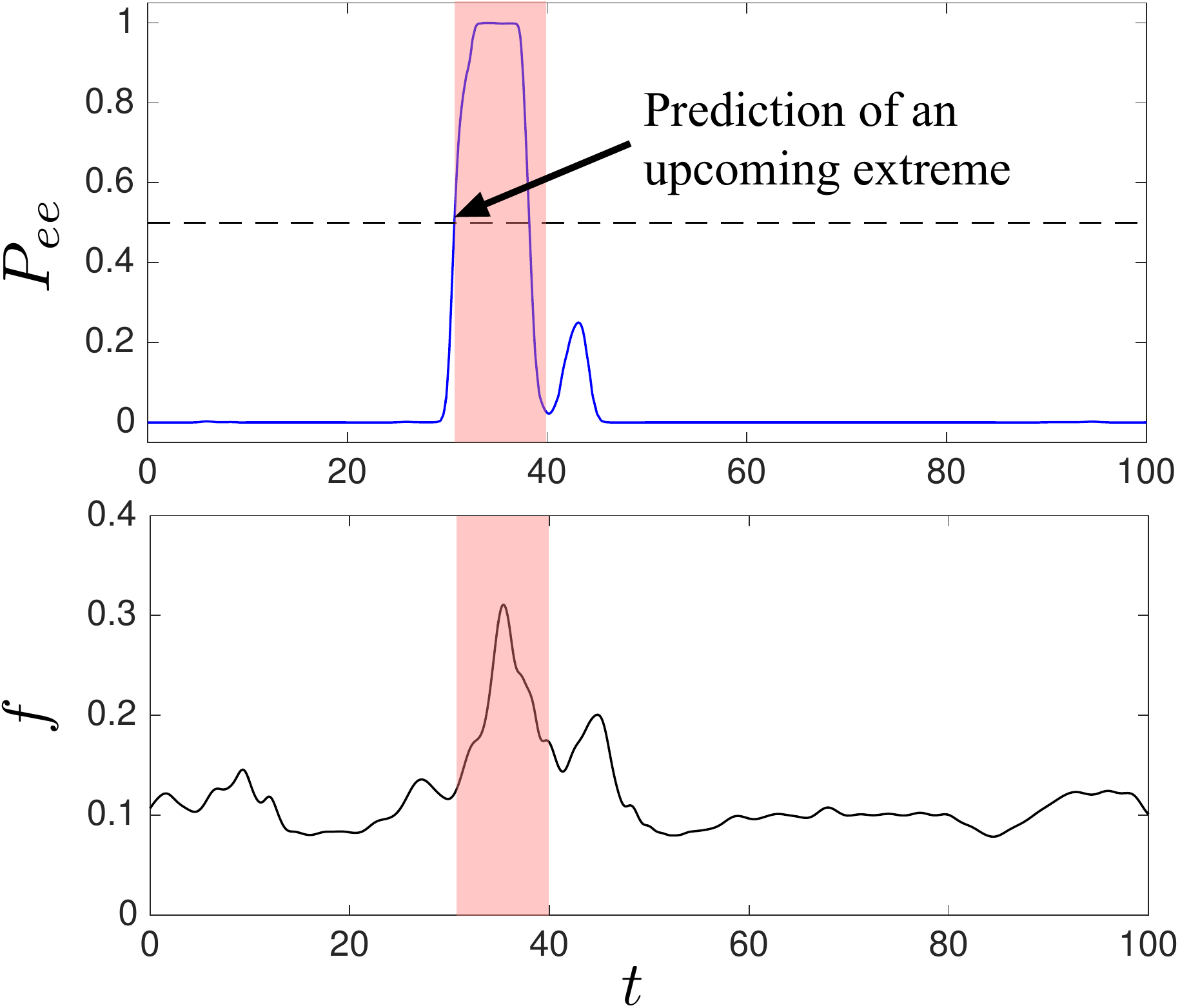}}
\caption{Prediction of extreme events in the Kolmogorov flow at Reynolds number $Re=40$
and forcing wave number $k_f=4$.
(a) Conditional probability density $\subs{p}{f_m|g}$ where $f_m(u;t_1,t_2)=\max_{\tau\in[t_1,t_2]}D(u(\tau))$ 
is the maximum future values of the energy dissipation rate to be predicted and the indicator $g(u)=-|a(1,0)|$ is the indicator.
(b) Probability of future extreme events $P_{ee}$ as a function of the indicator $g(u)=-|a(1,0)|$.
(c) An instance of an extreme event and its short-term prediction signaled by $P_{ee}=0.5$. The observable 
being predicted is the energy dissipation rate, i.e., $f(u(t)) = D(u(t))$.
}
\label{fig:kolm_pred}
\end{figure*}

The conditional statistics discussed above allows us to quantify the extent to which such predictions are
feasible. Figure~\ref{fig:kolm_pred}(a) shows the
conditional PDF $\subs{p}{f_m|g}$ where the observable $\obs$ is the energy dissipation rate~\eqref{eq:ID}, i.e,
$\obs(u)=D(u)$. The indicator is chosen to be $g=-|a(1,0)|$. The minus sign ensures that relatively
large values (although negative) of the indicator correlate with the large values of the observable.

We point out a number of the important features of this figure. Most importantly, very small portion of 
the conditional probability density resides in the false positive or false negative regions (quadrants III and IV).
Since the extreme events are rare, most of the density is concentrated in $\obs_m<\obs_e$ region. This region
correlates strongly with $g<g_e$ which means the indicator successfully rules out
the non-extreme dynamics (quadrant I). Conversely, we see also a high correlation between $\obs_m>\obs_e$ 
and $g>g_e$ which means that the indicator successfully identifies upcoming extreme events. 
This is better captured through the resulting probability of upcoming extreme events $P_{ee}$
shown in figure~\ref{fig:kolm_pred}(b). For $g<g_e=-0.5$, we have $P_{ee}\simeq 0$ that means 
the probability of upcoming extreme events is almost zero. Conversely, for $g_e>-0.3$, we have
$P_{ee}\simeq 1$, that is an extreme event is almost certainly upcoming. 
Due to the monotonicity of $P_{ee}$, there is a point where $P_{ee}=0.5$ corresponding 
to an indicator threshold $g=g_e$ which in this case is approximately $-0.39$.

\begin{figure*}
\centering
\includegraphics[width=.7\textwidth]{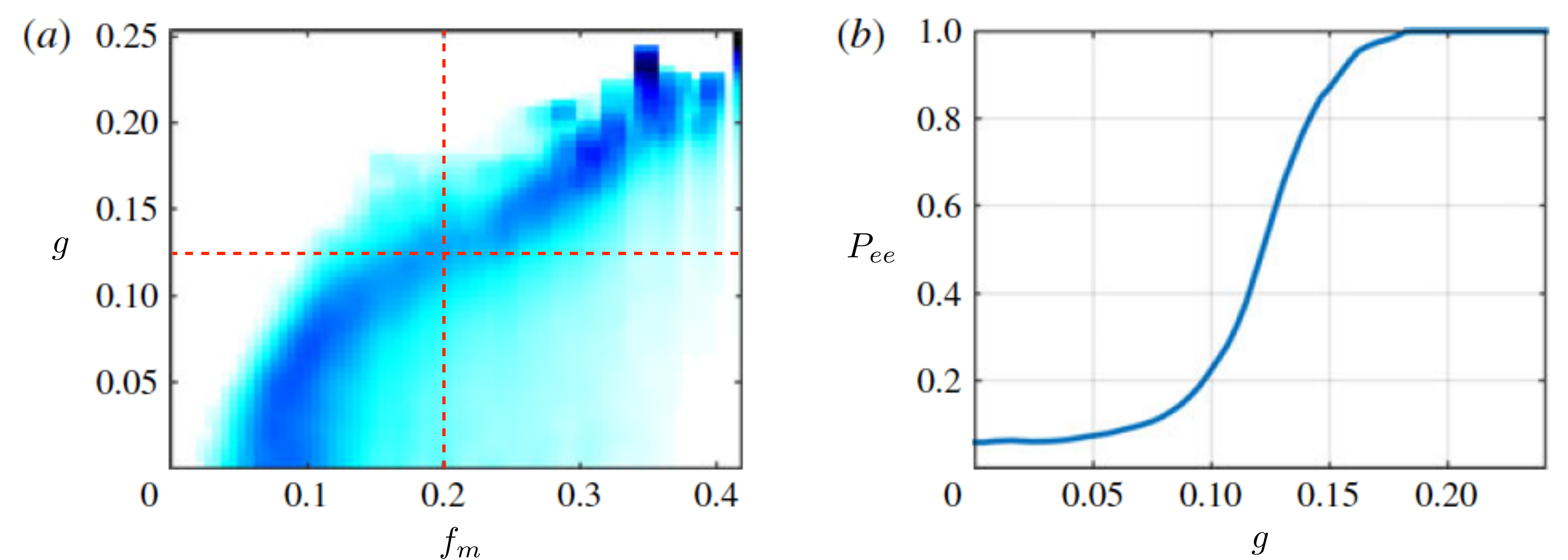}
\caption{Prediction of rogue waves. (a) The conditional PDF $\subs{p}{f_m|g}$
where the observable is the maximum wave height, $\obs(u) = \max_{x}|u(x,t)|$.
(b) The resulting probability of extreme events $P_{ee}$ as defined in Definition~\ref{def:Pee}.
Figure reproduced from~\cite{cousins16}.}
\label{fig:wave1d_pred}
\end{figure*}

Figure~\ref{fig:kolm_pred}(c) shows the application of the indicator to predicting extreme events
along a trajectory of the Kolmogorov flow. Along this trajectory, the indicator $g$ is measured 
and the resulting $P_{ee}(g)$ is computed. Most of the time, the probability of extreme events
is almost zero. At around time $t=30$, however, this probability increases rapidly and 
eventually passes the threshold $P_{ee}=0.5$, signaling an imminent extreme event in the near future. 

Similar results are obtained for the prediction of rogue waves.
Recall from section~\ref{sec:wave}  that rogue waves develop from
localized wave groups with certain range of length scales and amplitudes.
Cousins and Sapsis~\cite{cousins16} proposed an indicator $g$ of upcoming rogue waves
by projecting the wave envelope $u$ unto a subspace which captures this `dangerous'
range of length scales and amplitudes (see Ref.~\cite{cousins16} for further details). 
The larger the projection, the more likely is the occurrence of a future rogue wave. 

Figure~\ref{fig:wave1d_pred} shows the resulting conditional PDF $\subs{p}{\obs_m|g}$ and the probability of upcoming
rogue waves $P_{ee}$. Here, the observable is the maximum wave amplitude over the entire domain, i.e.,
$\obs = \max_x|u(x,t)|$. This conditional PDF has a similar structure to that of the Kolmogorov flow shown 
in figure~\ref{fig:kolm_pred}: strong correlation between small (resp. large) values of the indicator $g$
and the small (resp. large) values of the future observable $\obs_m$.
However, the conditional probability density in figure~\ref{fig:wave1d_pred}(a) has a more significant 
density in quadrant III, i.e., there is a higher probability of false negatives. 
This is also reflected in the probability of upcoming extremes $P_{ee}$ shown in figure~\ref{fig:wave1d_pred}(b).
Note that even for small values of the indicator, $g<0.1$, there is a non-negligible probability of extremes, $0.05<P_{ee}<0.2$.
Contrast this with figure~\ref{fig:kolm_pred}(b) where for small indicator values, the probability of future extremes is almost zero.

Nonetheless, the false negatives comprise only $5.9\%$ of the predictions which is relatively low. 
A more reliable indicator of future rogue waves would have an even lower rate of false negatives (as well as false positives).
In the next section, we discuss possible methods for discovering most reliable indicators of extreme events for a given dynamical system.

\subsection{Data-driven discovery of indicators}\label{sec:pred_datadriven}
In section~\ref{sec:pred_app}, we demonstrated that the analysis of the structure of the governing equations
assisted with the variational method of section~\ref{sec:probe} can lead to the discovery of reliable indicators
of extreme events. In the Kolmogorov flow (section~\ref{sec:kolm}), for instance, 
we showed that such a reliable indicator is the modulus of a particular Fourier mode. 

This approach relies on the
solution of a solutions of a constrained optimization problem involving the governing equations of the system. 
One may wonder whether there is a purely data-driven method for discovery of reliable indicators of extreme events.
For Kolmogorov flow, for instance, it is quite possible that a carefully 
customized data analysis technique, applied to a long term simulation data,
could have led to the discovery of the same indicator. 

To date, a systematic framework for discovery of indicators of extreme events from data is missing. 
In the remainder of this section, we briefly sketch properties that such an approach should have.
Recall that a reliable indicator of extreme events should return low rates of false positive and false negative predictions. 
An indicator that constantly issues alarms of upcoming extremes will correctly `predict' the extreme events. However,
this indicator is not desirable since it also returns a large number of false alarms. 
Conversely, an indicator that never issues an alarm, will have no false alarms but 
will also miss all the extreme events. 
Therefore, a reliable indicator is one that returns minimal number of combined false positives and
false negatives.

The false positive and false negative predictions can be combined into a quantity called the \emph{failure rate}.
For an observable $\obs:\funcspace\to\mathbb R$ with the extreme event threshold $\obs_e\in\mathbb R$,
the failure rate of an indicator $g:\funcspace\to\mathbb R$ is
\begin{align}
\mathcal L(g;g_e,t_1,t_2) :=& 
\Pmeas\left( u\in\funcspace: \obs_m(u;t_1,t_2)>f_e | g(u)<g_e\right)+ \nonumber\\
& \Pmeas\left( u\in\funcspace: \obs_m(u;t_1,t_2)<f_e | g(u)>g_e\right),
\label{eq:FR_00}
\end{align}
where $\obs_m(\cdot;t_1,t_2):\funcspace\to\mathbb R$ is the future maximum of the observable $\obs$ as defined in~\eqref{eq:fm}
and $g_e\in\mathbb R$ is the alarm threshold such that $g>g_e$ signals an upcoming extreme event.
Note that $\mathcal L(\cdot;g_e,t_1,t_2) : L^\infty (\funcspace)\to [0,1]$
measures the probability of false negative ($\obs_m>\obs_e$ given that $g<g_e$)
and false positive ($\obs_m<\obs_e$ given that $g>g_e$) predictions.

It follows from the definition of the conditional PDF $\subs{p}{\obs_m|g}$ that the failure rate is equal to
\begin{align}
\mathcal L(g;g_e,t_1,t_2)=&\underbrace{\int_{f_e}^\infty\int_{-\infty}^{g_e} \subs{p}{f_m|g}(a,b)\id b\,\id a}_\text{False Negatives}+\nonumber\\
&\underbrace{\int_{-\infty}^{f_e}\int_{g_e}^{\infty} \subs{p}{f_m|g}(a,b)\id b\,\id a}_\text{False Positives}.
\label{eq:FR_01}
\end{align}
This expression measures the conditional density residing in quadrants (III) and (IV) of figure~\ref{fig:condPDF},
measuring the false negatives and the false positives, respectively. 

The failure rate depends on the indicator $g\in L^\infty(\funcspace)$ and three parameters, $g_e$, $t_1$ and $t_2$. 
The main objective is to find an indicator $g$ that minimizes the failure rate. However, one should
simultaneously search for the appropriate parameters $(g_e,t_1,t_2)$. 
In figure~\ref{fig:kolm_pred}, for instance, the predictions correspond to $t_1=1$ and $t_2=t_1+1$.
If we gradually increase the prediction horizon $t_1$ the prediction skill of the indicator deteriorates such that for $t_1>4$
the indicator loses any predictive power. This finite-time predictability is expected since in chaotic systems
the observables tend to have finite correlation times. A similar observation is valid for the indicator threshold $g_e$.
Therefore, the minimization of the failure rate $\mathcal L$
should be carried out simultaneously over the measurable observables $g$ and the parameters $(g_e,t_1,t_2)$.

The resulting minimizer is a reliable indicator of extreme events. Solving this optimization problem, however,
is not straightforward because of the nonlinear and non-smooth dependence of the failure rate $\mathcal L$
on the function $g$ (see Eqs.~\eqref{eq:condPDF} and~\eqref{eq:FR_01}). 
Treatment of this optimization problem will be addressed elsewhere.

\section{Summary and conclusions}\label{sec:concl}
The study of extreme events can be divided into four components: mechanisms, real-time prediction, 
mitigation and statistics. Compared to the last topic  that has been studied deeply for certain systems~\cite{varadhan2008,davison2015}, 
the other three aspects have received less attention. In this review, we focused on 
two of these aspects, namely mechanisms and real-time prediction, and reviewed the quantitative treatment of them. 

Mechanisms that lead to the formation of extreme events are not unique. Depending on the system, they can be, for instance, a result of
multiscale instabilities, driven by noise or the consequence of nonlinear energy transfers. Yet, our review 
suggests that there might be a unified mathematical framework for discovering these mechanisms. 

In high-dimensional chaotic systems, the mechanisms underlying the extreme events are usually difficult to discern by relying solely on observation (or simulation) data.
A successful method for discovering the underlying mechanisms should take a blended approach combining the governing equations of the system
with the observation data or some low-order statistics. For instance, the variational method of Section~\ref{sec:probe} seeks the extreme event mechanisms as the
solutions of a constrained optimization problem. Here the governing equations are used to form an appropriate objective functional
and the observation data is used to form the appropriate constraints.

Prediction of individual extreme events is another aspect reviewed here.
The prediction problem consists of designing a reliable indicator function
whose behavior (e.g., large values) signals an upcoming extreme event. 
A reliable indicator is one that returns relatively low rates of 
false positive and false negative predictions. We saw that even partial knowledge of the mechanisms that lead to the extremes 
can inform the choice of a reliable indicator. While the discovery of the mechanisms
relies on the governing equations, the predictions can be performed
in a purely data-driven fashion. This of course assumes that the derived indicator can 
be measured or observed in practice. 

Discovery of reliable indicators of extreme events directly from observed data is highly desirable. 
This is specially the case for problems, such as earthquakes, epileptic seizures, and social dynamics, where the governing equations are unknown. 
In section~\ref{sec:pred_datadriven}, we sketched some desirable properties that such a 
reliable indicator should have. We also outlined several technical problems surrounding this approach that remain unresolved
and should be addressed in future work.

As for the mitigation of extreme events, very little has been done. The existing studies
are narrow in scope and revolve around arbitrary perturbations that may nudge the 
system away from extreme events. Therefore, a control theoretic study of 
the mitigation of extreme events merits further investigation. This includes formulating the problem
in terms of observable quantities that can be measured in practice, as well as control variables that can be
adjusted. This greatly limits the admissible perturbations to the system and sheds light on the practical limitations
of mitigating extreme events.

Finally, we point out that our discussion of extreme events was limited to autonomous systems. 
These are systems governed by a fixed set of principles. Our discussion does not apply to non-autonomous systems, such as stock markets~\cite{drazen2000}
or social networks~\cite{keizer2008,rand2011}, where the rules of engagement can change over time. While non-autonomous dynamical systems
have been studied extensively~\cite{kloeden2011,carvalho2012}, the literature on extreme events in these systems is vanishingly small and remains an
attractive area to be investigated.

\begin{acknowledgment}
This work has been supported through the ARO MURI grant W911NF-17-1-0306, the ONR MURI grant N00014-17-1-2676 and the AFOSR grant FA9550-16-1-0231.  
\end{acknowledgment}

%
% ...

%

%

\end{document}